# Quadripartite bond length rule applied to two prototypical aromatic and antiaromatic molecules

Łukasz Wolański*[a], and Wojciech Grochala*[a]

This work is dedicated to Professor Roald Hoffmann at his 85[th] birthday and it also commemorates the resonance (partial valence bond) theory of benzene molecule by Friedrich Karl Johannes Thiele

[a]  Ł. Wolański, W. Grochala
Centre of New Technologies, University of Warsaw,
S. Banacha 2c, 02-097 Warsaw, Poland
[*]  E-mail: l.wolanski@cent.uw.edu.pl, w.grochala@cent.uw.edu.pl

**Abstract:** In 2000, a remarkably simple relationship was introduced, which connected the calculated geometries of isomolecular states of three different multiplicities. These encompass a ground singlet state, the first excited triplet state, as well as related radical anion and radical cation. The rule allows prediction of geometry of one of the species if the three remaining ones are known. Here, we verify applicability of this bond length rule for two small planar cyclic organic molecules, i.e. benzene and cyclobutadiene, which stand as prototypical examples of, respectively, aromatic and antiaromatic systems. We see that the rule works fairly well to benzene and it works independently for quinoid as well as for anti-quinoid minima, and despite the fact that radical anion species poses challenges for correct theoretical description.

## Introduction

Geometry optimization as a multi-step process is usually the longest part of the typical quantum chemical computations. For that reason, geometry optimization is often carried out using a less precise method than the final technique used for determining other properties. Alternatively, a pre-optimization may be carried out at some low level of theory, followed by a more rigorous one, particularly for large molecules. Geometry optimization is also a process that requires some chemical knowledge and intuition, since an initial atoms positions have to be defined first. An unrealistic choice of a starting geometry can lead to their convergence to the structures corresponding to saddle points and/or calculations may take an impractically long time. This is particularly important when it concerns the optimization of molecules in their excited electronic states, or of the even-electron (free radical) systems. In principle, this additionally requires the use of more resource-consuming methods (i.e. CC2 vs. MP2, TDDFT vs. DFT, with rather multi- than single-configurational wavefunction) than in the case of the ground-state equilibrium structures.

For these reasons, it is highly desirable to be able to propose a starting structure for optimization which is a result of an educated guess and as such could lead to much faster convergence. About two decades ago, Grochala, Albrecht, and Hoffmann have observed a remarkably simple relationship (subsequently labelled by Parr and Ayers as "GAH rule") *i.e.* the corresponded bond lengths in the cationic ($R^+$), anionic ($R^-$) and neutral ($R^0$) systems (all in their electronic ground states) together with neutral structure in its first triplet excited state ($R^0_{T1}$) approximately satisfy the equation (Eq.1):

$$\Delta GAH(R) = R^+ + R^- - R^0_{T1} - R^0 \approx 0 \qquad (Eq.1)$$

This approximate relationship was proposed based on rather low-level (at least by today standards) quantum mechanical calculations for a handful of inorganic (largely diatomics) and organic molecules, both neutral, cationic and anionic, among those $C_2$, $C_2H_2$, $C_2H_4$, $C_2H_6$, $N_2H_2$, $B_2H_2$, CO, $CN^-$, $N_2$, $NO^+$, and three more complex hydrocarbons. These authors have noted that relationship expressed by Eq.1 seems to be most accurate, if the ground state molecule is nondegenerate and equilibrium geometries of all structures ale reasonable similar. [1] Moreover, it applies exclusively to bonds constituting the chromophore part of a molecule and works best for systems with conjugated double bonds.

While computational effort related to verifying the validity of the GAH rule was rather limited, the reasons behind its seeming success are far from being obvious. Indeed, the quest for the rule was inspired by a simplistic molecular orbital (MO) picture, and perturbation theory in its most simple implementation, which may obviously be of a great didactic value. *I.e.*, if one uses a one-orbital basis set for each atom in a diatomic, a classical two-MO picture of electronic structure emerges, with the bonding Highest Occupied Molecular Orbital (HOMO) and antibonding Lowest Unoccupied Molecular Orbital (LUMO) orbital. The two-electron singlet ground state has HOMO doubly occupied and an empty LUMO, so it maximizes the bonding between both atoms. A subtraction of one electron (to form a radical cation) may be treated as a perturbation, due to which bonding strength decreases and interatomic bond elongates. An addition of one electron (to form a radical anion) is yet another perturbation, due to which the antibonding effect appears and the bond weakens and elongates again. Usually, the bond weakening associated with removal of one electron from HOMO is slightly weaker than the one related to occupation of LUMO by one electron, and that is because "the antibonding orbital is more antibonding than the bonding orbital is bonding". Nevertheless, the formation of an excited triplet state from the ground state singlet is associated with *both* effects in the same time, i.e. with two effectively antibonding effects due to decreased occupancy of HOMO and increased one of LUMO simultaneously. Therefore, it is not totally unexpected that the bond weakening is now more-less a sum of the two effects seen for two distinct free radical species.

Based on the MO theory in its two-center two-orbital implementation, one may additionally deduce that the GAH rule should work best for electronic manifold made up from π type orbitals rather than σ ones. This is because the first excited triplet state of a σ bond corresponds usually to a fully dissociated bond, and thus the effects of the singlet → triplet excitation for molecular geometry are so large that such excitation cannot be treated as a small perturbation of a system. This naturally explains why the applicability of the rule was documented before for systems with double or triple bonds, either isolated or conjugated. It is also easy to understand why the rule finds most applicability for the chromophore part of a molecule; note that a bond very distant from a chromophore and not conjugated with it via a π system does not experience any major bond length changes upon one-electron perturbations within the chromophore (i.e., for such distant bond, $R^+ = R^- = R^0 = R^0_{T1}$) and therefore the relationship expressed by Eq.1 still holds but it becomes trivial.

While the MO theory served as initial inspiration of the rule, its applicability to diverse molecular systems is far from ideal, especially when strong electronic correlation effects apply. In a separate line of reasoning, Ayers and Parr managed to show, how the Fukui function can rationalize this rule by noticing that for nondegenerate ground states it is advantageous to have a large band gap – i.e. chemically hard systems should best follow the rule.[2] Following that, an extension of 'GAH approximation' was proposed by Morell and co-workers as a way of calculating the potential energy profile of reaction in its first electronic excited state.[3]

With the enormous advances of supercomputing power which took place during the last quarter of a century, it was tempting for us to check the validity of the GAH rule using higher-level computational methods. We have selected two molecules for this study, i.e. cyclobutadiene (CBDE) and benzene (BZ). First, these molecules stand for prototypical antiaromatic (4e) and aromatic (6e) systems, respectively. On the other hand, their first excited triplet states are aromatic (2e) and antiaromatic (4e), respectively. In other words, the perturbation associated with one electron singlet → triplet excitation flips the aromaticity entirely to its opposite, a true revolution in electronic properties. Secondly, both molecules are relatively small, which permits quantum mechanical calculations to be performed at quite precise reference computational levels. Third, having large HOMO/LUMO gap these small systems seem to be ideally suited for such test, according to Ayers and Parr predictions[2]. Fourth, both systems are cyclic which introduces a certain constrain on their geometry due to persistent σ bond framework. And last but not the least, BZ in its two radical ion forms as well as in the excited triplet state offers *two* distinct minima to be independently studied; one corresponds to a quinoid type (with two short and four long C–C bonds) and another to anti-quinoid one (with four short and two long bonds), so each of those may looked at separately (Figure 1). Despite their small size, the two molecules selected for this study host multiple fascinating phenomena and constitute a playground for theoretical methods. Take benzene; with only 6 π orbitals, not just two Kekulé ones and three Dewar ones but as many as 175 well-defined covalent and ionic valence bond structures are possible.[4] This leads to a multiconfigurational character of the this molecule of immense complexity and despite its seemingly simple regular geometry. The delocalized π-electron component of benzene is stabilized by resonance, but is also destabilized by localizing distortions, what is unfortunately a much less acknowledged fact.[5] Ulusoy and Nest shown that the aromaticity of benzene in its electronic ground state can simply be switched off by an ultrashort laser pulse.[6] The antiaromatic triplet state of benzene also exhibits many unusual complexities [7–11]. Clearly, there is still a lot to be learned from these small molecules.

## Results and Discussion

We decided to utilize three types of computational approaches: fast and efficient density functional theory (DFT), the coupled cluster (CC) method, complete active space self-consistent field (CASSCF) approach, and the most accurate but also resource-consuming perturbation theory with multireference wavefunction (CASPT2). Calculations for all these methods were performed with at least three different basis sets, as described in **Computational Details** section. The symmetry was constrained to $D_{6h}$ for singlet BZ, $D_{4h}$ for triplet CBDE, and $D_{2h}$ for all other species. Thus, in each molecule there are at most two distinct C-C bond lengths labelled as **a** and **b** (Figure 1). Each of those may in principle take different values for the ground singlet state in a neutral molecule, its first excited triplet state, as well as doublet states of the radical anion and radical cation, and they are labelled here as $a^0$ or $b^0$, $a^0_{T1}$ or $b^0_{T1}$, $a^-$ or $b^-$, and $a^+$ or $b^+$, respectively. The calculated bond lengths are collected in Table 1 (only for the largest basis sets studied here for each given level of theory, usually cc-pVQZ) and in Tables S1-S14 in Supplementary Information (an extended set for all basis sets studied).

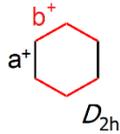

**Figure 1.** Molecular symmetry and C-C bond length labelling for investigated electronic states in quinoid (Q) and anti-quinoid (AQ) isomers of benzene (BZ) species, as well as in cyclobutadiene (CBDE).

We begin by noticing that the predicted C–C bond lengths for each species separately are quite dependent mainly on the computational method (*cf.* SI). While, the C-C bond length in singlet BZ, $a^0$, varies between 1.385 Å and 1.409 Å (*i.e.* by 0.024 Å), the discrepancy for its triplet state is much larger (0.049 Å for bond length $a^0_{T1}$). For CBDE, the smallest and the largest discrepancies are observed for radical anion (0.025 Å for bond length $a^-$) and neutral singlet form (0.078 Å for $b^0$), respectively.

**Table 1.** C-C bonds lengths [Å] in quinoid (Q) and anti-quinoid (AQ) conformers of benzene (BZ) and in cyclobutadiene (CBDE). For bonds labelling for molecules in investigated electronic states see **Fig. 1**. ΔGAH(**R**) value [Å] was calculated according **Eq. 1**. Computational data for the most resource-demanding CASPT2 approach are for cc-pVTZ basis set (*italics*), whereas all others are for cc-pVQZ. All values are rounded to three decimal places. Results obtained for other investigated basis sets are available in ESI. max(**R**) – min(**R**), where R=**a** or **b**, denotes the span of the bond length values between all methods tested for each molecular species and each bond separately.

|  | Computational approach | a$^+$ | a$^-$ | a$^0$$_{T1}$ | a$^0$ | ΔGAH(a) | b$^+$ | b$^-$ | b$^0$$_{T1}$ | b$^0$ | ΔGAH(b) |
|---|---|---|---|---|---|---|---|---|---|---|---|
| **BZ(AQ)** | *CASPT2(ε=0.25)* | *1.447* | *1.456* | *1.499* | *1.393* | ***0.011*** | *1.387* | *1.395* | *1.392* | *1.393* | ***-0.003*** |
|  | CASSCF | 1.440 | 1.452 | 1.499 | 1.392 | **0.001** | 1.383 | 1.392 | 1.391 | 1.392 | **-0.008** |
|  | CC2 | 1.441 | 1.443 | 1.503 | 1.394 | **-0.012** | 1.370 | 1.395 | 1.390 | 1.394 | **-0.018** |
|  | DFT(B3LYP) | 1.447 | 1.456 | 1.517 | 1.391 | **-0.004** | 1.383 | 1.393 | 1.382 | 1.391 | **0.003** |
|  | DFT(M06-2X) | 1.446 | 1.452 | 1.514 | 1.388 | **-0.003** | 1.380 | 1.389 | 1.379 | 1.388 | **0.002** |
|  | DFT(CAM-B3LYP) | 1.443 | 1.451 | 1.513 | 1.385 | **-0.005** | 1.377 | 1.386 | 1.377 | 1.385 | **0.001** |
|  | max(**R**) – min(**R**) | 0.007 | 0.013 | 0.018 | 0.009 | --- | 0.017 | 0.009 | 0.015 | 0.009 | --- |
| **BZ(Q)** | *CASPT2(ε=0.25)* | *1.370* | *1.378* | *1.358* | *1.393* | ***-0.003*** | *1.426* | *1.435* | *1.463* | *1.393* | ***0.005*** |
|  | CASSCF | 1.364 | 1.373 | 1.353 | 1.392 | **-0.008** | 1.421 | 1.431 | 1.466 | 1.392 | **-0.004** |
|  | CC2 | 1.356 | 1.380 | 1.355 | 1.394 | **-0.013** | 1.412 | 1.432 | 1.464 | 1.394 | **-0.014** |
|  | DFT(B3LYP) | 1.364 | 1.374 | 1.341 | 1.391 | **0.007** | 1.425 | 1.434 | 1.470 | 1.391 | **-0.002** |
|  | DFT(M06-2X) | 1.360 | 1.370 | 1.335 | 1.388 | **0.007** | 1.424 | 1.430 | 1.470 | 1.388 | **-0.003** |
|  | DFT(CAM-B3LYP) | 1.357 | 1.367 | 1.333 | 1.385 | **0.006** | 1.420 | 1.428 | 1.467 | 1.385 | **-0.003** |
|  | max(**R**) – min(**R**) | 0.014 | 0.013 | 0.025 | 0.009 | --- | 0.014 | 0.007 | 0.007 | 0.009 | --- |
| **CBDE** | *CASPT2(ε=0.25)* | *1.500* | *1.512* | *1.438* | *1.557* | ***0.017*** | *1.380* | *1.398* | *1.438* | *1.350* | ***-0.010*** |
|  | CASSCF | 1.482 | 1.499 | 1.435 | 1.547 | **-0.001** | 1.374 | 1.390 | 1.435 | 1.346 | **-0.017** |
|  | CC2 | 1.492 | 1.505 | 1.435 | 1.561 | **0.002** | 1.378 | 1.395 | 1.435 | 1.340 | **-0.001** |
|  | DFT(B3LYP) | 1.500 | 1.512 | 1.436 | 1.574 | **0.002** | 1.373 | 1.390 | 1.436 | 1.329 | **-0.002** |
|  | DFT(M06-2X) | 1.494 | 1.505 | 1.430 | 1.566 | **0.003** | 1.369 | 1.385 | 1.430 | 1.325 | **-0.001** |
|  | DFT(CAM-B3LYP) | 1.493 | 1.504 | 1.429 | 1.566 | **0.003** | 1.367 | 1.384 | 1.429 | 1.323 | **-0.001** |
|  | max(**R**) – min(**R**) | 0.018 | 0.013 | 0.009 | 0.027 | --- | 0.013 | 0.014 | 0.009 | 0.027 | --- |

Obviously, in our analysis we will seek validation of Eq.1 for one given level of theory at a time.

It is quite disturbing that the diverse computational methods often disagree on whether a planar geometry studied here is a local minimum, or not. For high symmetry systems with "simple" electronic structure (triplet and singlet CBDE and singlet BZ), we find nearly no discrepancies, as expected. Still, ground state benzene singlet turns out to show one imaginary frequency at CC2/aug-cc-pVTZ level, and quite appreciable one (i423 cm$^{-1}$) (*sic!*). The computational methods consistently yield local minimum nature of the Q form of BZ radical cation, and for the CBDE radical cation. Yet for AQ form of BZ radical cation we find that all DFT(M06-2X) calculations predict a transition state nature for this planar species, in contrast to all remaining methods.[12,13] But the remaining open-shell species yield a true diversity of the number of imaginary frequencies. For example, the problematic BZ radical anion in its Q form is either a local minimum (according to selected DFT approaches), a transition state (selected CASSCF and DFT results), or a saddle point of the second order (according to the remaining CASSCF and DFT approaches). Similar problems are also encountered for BZ triplet state (mostly in the Q form). What is even more puzzling, though, is that the CBDE radical anion (which, as a 3-electron 4-center system *must* host the Jahn-Teller effect in $D_{4h}$ geometry) is computed by some DFT methods (*e.g.* CAM-B3LYP with either 6-31G(d,p) or cc-pVDZ basis set) to be an undistorted square. These problems are not unprecedented.[12] The sensitivity of the potential energy surface details to the level of theory for these species is fascinating[14] and certainly comes from the fact that for relatively small 4 and 6 electron systems any 1e perturbation leading to a free radical may be considered large. The discrepancies between various methods documents the problems which a theoretician faces while trying to correctly predict the geometry of a ground state. Fortunately, our task to verify the validity of the GAH rule for the pure π manifold (*i.e.* free from admixture of σ one) permits us to comfortably neglect these issues by restricting ourselves to planar geometries, independent of their character (*i.e.* a true minimum or not). Moreover, as wave-function based methods are usually more sensitive than DFT ones to the choice of a basis set, from now on we discuss hereonly results obtained with the largest basis set (cc-pVQZ, except for the most resource-demanding CASPT2 where cc-pVTZ was used). For the remaining ones *cf.* the SI.

**Cyclobutadiene – antiaromatic system.** Let us begin with the smaller and less complex of two systems, *i.e.* CBDE. The molecular orbital (MO) system of the π manifold for an ideal square $D_{4h}$ symmetry is shown in Figure 2 (left). The lowest energy orbital is nondegenerate, and it is bonding between all pairs of C atoms. Its fully antibonding equivalent is also nondegenerate, and obviously it has the smallest binding energy.

On the other hand, there are two different but energy-equivalent combinations of atomic orbitals, which are bonding between two pairs of C atoms and antibonding between the remaining two pairs. The triplet state of CBDE corresponds to single occupation of each of the two degenerate orbitals and as such preserves $D_{4h}$ symmetry. However, an occupation of these two orbitals by either one electron (in a radical cation), three electrons (in a radical anion) or one electron pair (in a singlet state) corresponds to the Jahn-Teller scenario and it must lead to deformation of a square to a rectangle ($D_{2h}$). This removes the orbital degeneracy (Figure 2 right). Thus, a singlet ground state of CBDE is a prototypical antiaromatic system which exhibits bond alternation. This behaviour may be formulated in terms of the Maximum Hardness Principle[15–18]; the singlet state with one doubly filled and one empty degenerate orbitals would exhibit null electronic gap at the Fermi level, and infinite polarizability. The rectangular distortion leads to increase of electronic hardness via band gap opening.

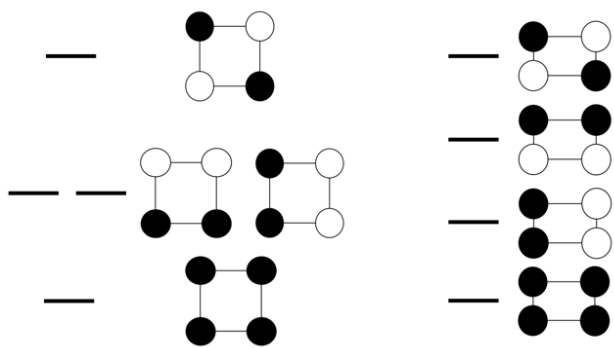

**Figure 2.** Illustration of the MO system of CBDE in $D_{4h}$ geometry (left) and in a lower $D_{2h}$ one (right). The rectangular deformation of a square has been exaggerated to facilitate detection of bond length changes.

Since the changes of molecular geometry (*i.e.* short-long C-C bond pattern) are interconnected with the bonding/antibonding character of the two frontier orbitals, one may expect the following inequalities to hold:

$\mathbf{a^0} > \mathbf{a^-} \approx \mathbf{a^+} > \mathbf{a^0_{T1}}$  (Eq.2) for **a** bond of CBDE
$\mathbf{b^0} < \mathbf{b^-} \approx \mathbf{b^+} < \mathbf{b^0_{T1}}$  (Eq.3) for **b** bond of CBDE

This is because double occupancy of an orbital which is antibonding between a given pair of C atoms leads to a larger lengthening of the bond than a single occupation of the same orbital, and this in turn leads to a larger lengthening of the bond than for the unoccupied case.

Indeed, we notice that according to CASPT2, CASSCF and CC2, the expected following inequalities consistently hold. However, all density-based approaches fail to show the expected bond length pattern for bond length **b**.

How well does the GAH rule hold for CBDE? It turns out that all methods wavefunction-based methods, ΔGAH(R) (Eq.1) does not exceed +/-0.017 Å. This is not a lot, given that these methods show discrepancies for individual bonds of a similar magnitude, *i.e.* up to 0.018 Å (Table 1). In one particular case, that of CC2 calculations, the ΔGAH(**a**) is as little as 0.002 Å, while ΔGAH(**b**) equals -0.001 Å. This implies that a 'hybrid species' constructed with the help of the GAH rule from the ground singlet as well as radical anion and radical cation, not only closely resembles an undistorted square, but also the CC-bond lengths of this hybrid fall extremely close to that calculated for the triplet state ($D_{4h}$).

**Benzene – aromatic system.** Let us now turn to a prototypical aromatic six-electron system – benzene. The MO scheme for BZ is illustrated in Figure 3. The π manifold for an ideal hexagonal $D_{6h}$ symmetry includes a totally-bonding nondegenerate orbital of the lowest energy, its nondegenerate totally-antibonding analogue of the highest energy, and two pairs of degenerate MO sets. For 6e occupancy in the ground singlet state the molecule shows an appreciable band gap among the two sets of the frontier orbitals, it is not subject to the Jahn-Teller effect, and it does not lower its symmetry. However, for either radical cation, radical anion or the lowest triplet state, the Jahn-Teller effect is operative, and the molecule distorts to $D_{2h}$. There are two local energy minima depending on the phase of the Jahn-Teller distortion, a quinoid one (Q) with two short and four longer CC bonds, and the antiquinoid one (AQ) with four short and two longer bonds. In such case, degeneracy of each previously doubly degenerate MO set is lifted. Obviously, this is also expected based on the Maximum Hardness Principle.

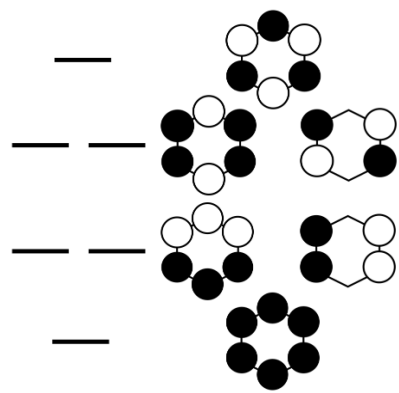

**Figure 3.** Illustration of the MO system of BZ in $D_{6h}$ geometry; in Q and AQ structures of the radical cation, radical anion, and of the triplet state, the symmetry is lowered to $D_{2h}$ (*cf.* Figure 1).

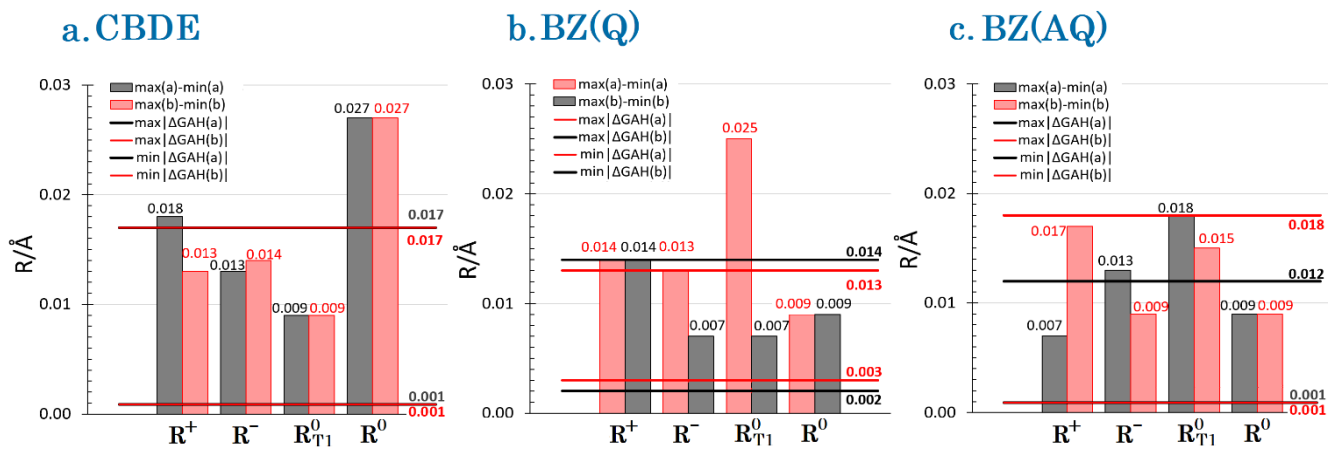

**Figure 4.** Minimum (**min|ΔGAH(R)|**) and maximum (**max|ΔGAH(R)|**) of absolute values of **Eq. 1** expression **max|ΔGAH(R)|** on the background of statistic ranges **max(R)-min(R)** of **a** and **b** bonds lengths of quinoid (**Q**) and anti-quinoid (**AQ**) benzene (**BZ**) conformers and for cyclobutadiene (**CBDE**) for all investigated computational approaches and only with the use of **the largest used basis sets** (cc-pVTZ for CASPT2 computations and cc-pVQZ for other approaches, for numerical data see **Table 1**). For statistics including results obtained with all basis sets see **Fig. S1**. Data for shorter bond are marked as red of pink, whereas data for longer ones are black or grey (see **Fig. 1**).

Here, the wavefunction as well as density-based approached yield quite similar results for all four species and for both Q and AQ minima separately. It turns out that the discrepancies between various methods in predicting bond lengths do not exceed 0.025 Å (for the case of **a** bond length of the triplet state in Q structure) but they are usually twice smaller and of the order of 0.01 Å (Table 1). In some cases, the discrepancies are even smaller than that. On the other hand, the GAH rule seems to perform reasonably well, with ΔGAH(R) of up to 0.018 Å for AQ structure and up to 0.014 Å for Q one. This result could be presented in the alternative way as follows: a 'hybrid species' constructed with the help of the GAH rule from the excited triplet as well as radical anion and radical cation, not only closely resembles an undistorted hexagon, but also the CC-bond lengths of this hybrid fall extremely close to that calculated for the ground singlet state ($D_{6h}$). This applies independently for the Q and for the AQ isomers. This result seems to suggest that the GAH rule may in principle hold for both aromatic (BZ) as well as antiaromatic systems (CBDE). A similar picture obtains when considering results obtained with the use of all methods as well as all basis sets applied here (*cf.* Figure S1 of the SI).

**Cyclopropene cation – an even smaller aromatic system.** While CBDE is a four-electron system, and thus represents the smallest antiaromatic system envisaged by the Hückel's rule, BZ with its six electrons is not the smallest aromatic system. Instead, systems such as cyclopropene cation (CP$^+$) in its ground singlet state provide the minimum number of two π electrons which may exhibit aromaticity. We will briefly discuss this case.

Now, the singlet state of CP$^+$ is already charged. Hence, removal of one electron leads to a radical dication, addition of one electron to a neutral radical, while the first excited triplet state of CP$^+$ is antiaromatic. To test applicability of the GAH rule for CP$^+$ we have used exclusively DFT methods. DFT is capable of correctly predicting an aromatic equilateral triangle ($D_{3h}$) geometry for the singlet ground state, with a relatively small discrepancy of the optimum bond length between various functionals (of up to 0.020 Å, *cf.* Table S17 in SI). The perturbation connected with adding or subtracting one electron leads to minima of the $C_{2h}$ symmetry, as expected. However, the destruction of the aromatic 2e π system upon excitation to the triplet state leads to a minimum which lacks $C_{2h}$ symmetry, and this does not permit to apply the GAH rule on the bond-by-bond basis. Moreover, one CC bond becomes so long (ca. 1.87-1.97 Å, *cf.* Table S18 in SI) that it may be considered broken. This means that the alternation of the 2e system of CP$^+$ is too large to treat it as a perturbation, since the sigma manifold is also greatly affected. In addition, Coulombic effects are huge for this small C3 molecule, especially for its dication radical, and they affect optimum geometry of diverse species used for derivation of the GAH rule to a greatly varying extent. Thus, and for many reasons, despite being a quite stable species, CP$^+$ is not a good testbed for verifying applicability of the GAH rule.

**GAH rule applied to spin densities.** The original work on the GAH rule has investigated applicability of the rule to all one-electron properties such as bond order etc. (*cf.* Appendix A and B[1]). Here, we were tempted to test applicability of the rule to atomic spin densities. Obviously, the rule applied to the total number of unpaired electrons, n, holds precisely, since:

ΔGAH(n) = 1(R$^+$) + 1(R$^−$) – 2 (R$^0_{T1}$) – 0 (R$^0$) = 0     Eq.2

However, it remains unclear whether the role would hold reasonably for local, i.e. atomic spin densities. In Tables 2 and 3 we list the atomic spin densities on C atoms according to the simplistic Mulliken population analysis. In Table 2, pure spin densities on C atoms are show, while in Table 3 densities of H atoms attached to each C atom has been summed up with those for the given C atom. It is observed that the 'residual' spin density, ΔGAH, calculated from an equation analogous to Eqns. 1 and 2, reaches the values very close to null for CBDE. The largest discrepancy from zero is -0.0026 e and it is computed for DFT(CAM-B3LYP) calculation. On the other hands, departures from zero are larger, but still not immense, *i.e.* up to -0.07 e, for BZ in both Q and AQ forms, if calculated with CASPT2 and CASSCF methods. DFT yields larger discrepancies up to -0.18 e for these species.

**Table 2.** Mulliken atomic spin densities for carbon atoms of investigated ring molecules obtained at chosen computational approaches. Because of molecular symmetry, we present here values for two unique carbon atoms in BZ (forming **a** and **b** bonds - marked as $C_{ab}$ and forming two **b** bonds - marked as $C_{bb}$, see Fig. 1) and for one in CBDE ($C_{ab}$). Values for ground electronic state $C(R^0)$ were predicted as combination $C(R^+) + C(R^-) - C(R^0_{T1})$ of atomic spin densities computed for other electronic states structures. Please note, that data for CASPT2 method are computed for equilibrium geometries obtained at this level of theory, however spin densities are computed basing on CASSCF wavefunction. All values are rounded to three decimal places and where necessary averaged because of molecular symmetry. Results obtained for other investigated basis sets are available in ESI.

| | Computational approach | $C_{ab}(R^+)$ | $C_{ab}(R^-)$ | $C_{ab}(R^0_{T1})$ | $C_{ab}(R^0)$ | $C_{bb}(R^+)$ | $C_{bb}(R^-)$ | $C_{bb}(R^0_{T1})$ | $C_{bb}(R^0)$ |
|---|---|---|---|---|---|---|---|---|---|
| **BZ(AQ)** | *CASPT2(ε=0.25)* | 0.2894 | | 0.5341 | | -0.0844 | | -0.0830 | |
| | CASSCF | 0.2868 | 0.2769 | 0.5316 | **0.0321** | -0.0803 | -0.0738 | -0.0826 | **-0.0715** |
| | DFT(B3LYP) | 0.2387 | 0.2331 | 0.6258 | **-0.1540** | 0.0136 | 0.0067 | -0.1972 | **0.2175** |
| | DFT(M06-2X) | 0.2389 | 0.2326 | 0.6017 | **-0.1302** | 0.0155 | 0.0143 | -0.2264 | **0.2562** |
| | DFT(CAM-B3LYP) | 0.2389 | 0.2325 | 0.6487 | **-0.1773** | 0.0140 | 0.0100 | -0.2387 | **0.2627** |
| **BZ(Q)** | *CASPT2(ε=0.25)* | 0.0396 | 0.0407 | 0.1106 | **-0.0303** | 0.4155 | 0.4073 | 0.7638 | **0.0590** |
| | CASSCF | 0.0414 | 0.0383 | 0.1053 | **-0.0256** | 0.4106 | 0.4029 | 0.7697 | **0.0438** |
| | DFT(B3LYP) | 0.0740 | 0.0681 | 0.1017 | **0.0404** | 0.3430 | 0.3368 | 0.8508 | **-0.1710** |
| | DFT(M06-2X) | 0.0766 | 0.0722 | 0.0800 | **0.0687** | 0.3403 | 0.3352 | 0.8071 | **-0.1317** |
| | DFT(CAM-B3LYP) | 0.0745 | 0.0694 | 0.0884 | **0.0555** | 0.3430 | 0.3361 | 0.8814 | **-0.2024** |
| **CBDE** | *CASPT2(ε=0.25)* | 0.2472 | 0.2451 | 0.4927 | **-0.0004** | | | | |
| | CASSCF | 0.2493 | 0.2489 | 0.4981 | **0.0001** | | | | |
| | DFT(B3LYP) | 0.2484 | 0.2478 | 0.4961 | **0.0001** | | | | |
| | DFT(M06-2X) | 0.2471 | 0.2450 | 0.4925 | **-0.0004** | | | | |
| | DFT(CAM-B3LYP) | 0.2466 | 0.2404 | 0.4896 | **-0.0026** | | | | |

**Table 3.** Mulliken atomic spin densities for carbon atoms (atomic spin densities of hydrogens are summed into C atoms they are connected to) of investigated ring molecules obtained at chosen computational approaches. Because of molecular symmetry, we present here values for two unique carbon atoms in BZ (forming **a** and **b** bonds - marked as $C_{ab}$ and forming two **b** bonds - marked as $C_{bb}$, see Fig. 1) and for one in CBDE ($C_{ab}$). Values for ground electronic state $C(R^0)$ were predicted as combination $C(R^+) + C(R^-) - C(R^0_{T1})$ of atomic spin densities computed for other electronic states structures. Please note, that data for CASPT2 method are computed for equilibrium geometries obtained at this level of theory, however spin densities are computed basing on CASSCF wavefunction. All values are rounded to three decimal places and where necessary averaged because of molecular symmetry. Results obtained for other investigated basis sets are available in ESI.

| | Computational approach | $C_{ab}(R^+)$ | $C_{ab}(R^-)$ | $C_{ab}(R^0_{T1})$ | $C_{ab}(R^0)$ | $C_{bb}(R^+)$ | $C_{bb}(R^-)$ | $C_{bb}(R^0_{T1})$ | $C_{bb}(R^0)$ |
|---|---|---|---|---|---|---|---|---|---|
| **BZ(AQ)** | *CASPT2(ε=0.25)* | 0.2925 | | 0.5419 | | -0.0852 | | -0.0838 | |
| | CASSCF | 0.2906 | 0.2873 | 0.5418 | 0.0361 | -0.0813 | -0.0745 | -0.0835 | -0.0723 |
| | DFT(B3LYP) | 0.2432 | 0.2465 | 0.5970 | -0.1073 | 0.0136 | 0.0069 | -0.1940 | 0.2145 |
| | DFT(M06-2X) | 0.2423 | 0.2428 | 0.6053 | -0.1203 | 0.0155 | 0.0145 | -0.2107 | 0.2406 |
| | DFT(CAM-B3LYP) | 0.2430 | 0.2449 | 0.6159 | -0.1280 | 0.0155 | 0.0102 | -0.2317 | 0.2559 |
| **BZ(Q)** | *CASPT2(ε=0.25)* | 0.0400 | 0.0419 | 0.1124 | -0.0305 | 0.4199 | 0.4162 | 0.7752 | 0.0609 |
| | CASSCF | 0.0420 | 0.0407 | 0.1076 | -0.0249 | 0.4161 | 0.4186 | 0.7847 | 0.0500 |
| | DFT(B3LYP) | 0.0753 | 0.0722 | 0.0926 | 0.0549 | 0.3494 | 0.3556 | 0.8148 | -0.1098 |
| | DFT(M06-2X) | 0.0776 | 0.0754 | 0.0927 | 0.0602 | 0.3449 | 0.3493 | 0.8145 | -0.1204 |
| | DFT(CAM-B3LYP) | 0.0756 | 0.0732 | 0.0804 | 0.0684 | 0.3488 | 0.3535 | 0.8391 | -0.1369 |
| **CBDE** | *CASPT2(ε=0.25)* | 0.2500 | 0.2500 | 0.5000 | 0.0000 | | | | |
| | CASSCF | 0.2500 | 0.2500 | 0.5000 | 0.0000 | | | | |
| | DFT(B3LYP) | 0.2500 | 0.2500 | 0.5000 | 0.0000 | | | | |
| | DFT(M06-2X) | 0.2500 | 0.2500 | 0.5000 | 0.0000 | | | | |
| | DFT(CAM-B3LYP) | 0.2500 | 0.2500 | 0.5000 | 0.0000 | | | | |

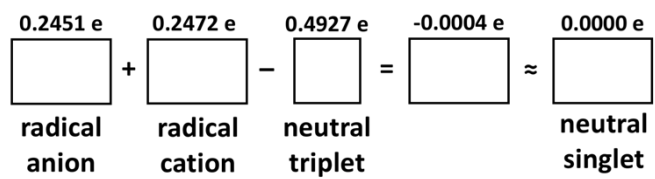

**Figure 5.** Illustration of the applicability of the GAH rule to atomic spin densities on C atoms for CBDE at the CASPT2 level (cf. Table 2). The rectangular deformation of a square has been exaggerated.

## Conclusions

We have reinvestigated the applicability of a simplistic bond length rule (GAH rule) for two cyclic molecules, CBDE and BZ, as prototypical examples of anti-aromatic and aromatic molecules, respectively, and using modern computational approaches of quantum chemistry. Rule was tested in its original spirit, *i.e.* assuming an enforced planar geometry of the carbon rings for all species considered. In general, the GAH rule is reasonably satisfied for these molecules, its errors not surpassing 0.018 Å for wavefunction-based methods and large basis sets. Moreover, the rule is reasonably fulfilled for atomic spin densities, particularly at CASSCF and CASPT2 levels, leading to rather small residual spin densities (typically +/-0.01 e and up to +/-0.07 e in some cases) for hybrids of radical anion, radical cation and triplet state, thus resembling a spin-less singlet state (Figure 5).

It is interesting that the rule seems to hold particularly due to the fact that the radical anion species is frozen here in the planar geometry, while it usually tries to break planarity for very small systems such as BZ or CBDE.[21,22] On the other hand, it is known that much larger aromatic hydrocarbons (or smaller, but fluorinated ones[19]) show positive electron affinity and they tend to form stable and planar radical anions; hence one might anticipate that the rule will apply very well for such systems. Yet, computational verification of the applicability of the rule for large systems requires more approximate quantum chemistry tools than those applied here, and this will be investigated in the near future.

## Computational Details

For CBDE, four species were studied: singlet ground state, the first excited triplet state, radical cation and radical anion. For BZ, each of the latter three were computed in two quinoid (Q) and anti-quinoid (AQ) form, thus leading to a total of seven distinct species.[20] In the spirit of the original paper,[1] we are looking at perturbations involving only the π manifold, free from any π–σ coupling.[21] Therefore, we perform geometry optimizations while constraining planarity of each system (more on that and related complexities in the Results and Discussion section).

As a reasonable compromise between computational cost and quality of results, coupled-cluster (CC) based methodology was initially utilized In this study. This computational approach was assed as very effective and accurate for theoretical researches of small and medium-sized organic molecules[22–24]. Therefore we decided to use the linear response approximate coupled-cluster of second order (CC2)[25,26] with efficient resolution-of-the-identity (RI) approximation [27,28] implemented in Turbomole 7.1 package[29,30]. CC2 computations were performed using Dunning and coworkers correlation-consistent basis sets (aug-)cc-pVXZ (X = D, T, Q) [31,32]. In case of CC2 computations, auxiliary basis sets[33] were also employed.

For comparison, all investigated structures were optimized also at (U)B3LYP[34–37], (U)M06-2X [38] and (U)CAM-B3LYP [39] density functional theory (DFT) levels of theory with Gaussian 16 revision A.03[40]. Dunning's cc-pVXZ (X = D, T, Q) [31,32] and Pople's 6-31G(d,p) basis sets were utilized.[41] *UltraFine* integration grid for numerical integrations and *Tight* geometry optimization criteria were used.

Finally, two multireference wave function based approaches were also utilized: complete active space self-consistent field method (CASSCF) [42–44] and the one based on its wavefunction second-order perturbation theory (CASPT2) [45,46]. Both type of calculations were carried with Molcas 8.0[47]. An active space was built with all π electrons and with 6 π-type orbitals, *i.e.* (6e, 6o) active space for neutral structures, (7e, 6o) for anionic and (5e, 6o) for cationic ones of BZ. Analogous active spaces for CBDE-based species also corresponded to all π electrons but now with 4 π-type orbitals. Considering the IPEA-shift parameter, [48,49] which modifies the zeroth-order Hamiltonian in CASPT2 method, geometry optimizations were performed here with either the present-day default value of this parameter (ε = 0.25 a.u., S-IPEA) [50] or without using it at all (ε = 0.00 a.u., 0-IPEA).

To test the character of obtained stationary points, and to cross check the validity of constraint on planarity of all species, vibrational frequencies were calculated with the same computational approach as for geometry optimization (except for all CASPT2 approaches and some of CASSCF ones, where computational cost exceeded resources available to us).

## Acknowledgements

W.G. gratefully acknowledges annual support from the Ministry of Education and Science. Calculations have been carried out using resources provided by Wroclaw Centre for Networking and Supercomputing (http://wcss.pl), grant No. 484.

**Keywords:** aromaticity • antiaromaticity • benzene • cyclobutadiene • molecular orbital theory

# Entry for the Table of Contents

Insert graphic for Table of Contents here.

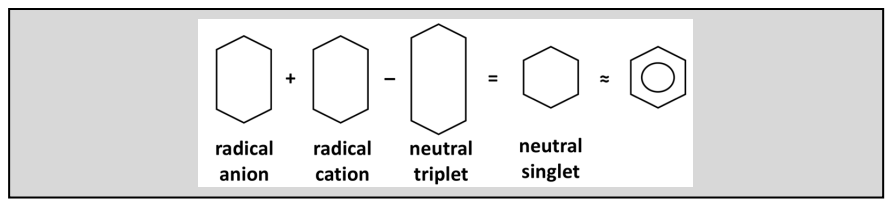

A geometrical hybrid of benzene radical cation, radical anion and the first excited triplet state of a neutral molecule adopts a nearly hexagonal symmetry and the bond length typical of the ground singlet state of benzene, as follows from the study of applicability of the GAH rule for prototypical aromatic and antiaromatic systems.

**SUPPORTING INFORMATION**

# Quadripartite bond length rule applied to two prototypical aromatic and antiaromatic molecules

*Łukasz Wolański\* and Wojciech Grochala\**

Centre of New Technologies, University of Warsaw, S. Banacha 2c, 02-097 Warsaw, Poland

\* e-mail: l.wolanski@cent.uw.edu.pl, w.grochala@cent.uw.edu.pl

# Table of contents

**Table S1.** Selected bond lengths [Å] between carbon atoms in quinoid-like anionic variant of benzene ring obtained from chosen approaches of quantum chemistry computational methods. C-C bond are labelled by designations introduced by **Figure 1**. If it is indicated, vibrational analysis was carried out for the equilibrium structure. All obtained structures are flat, possible imaginary frequencies relate to the swing of hydrogen atoms off the plane of symmetry, or are numerical artefacts of DFT methodology.

Total charge: -1
Spin multiplicity: 2
Geometry optimization state: $D_0$

**BZ(Q) R$^-$**

| Method | Basis set | Bond lenghts [Å] | | Imag. freq. |
|---|---|---|---|---|
| | | a$^-$ | b$^-$ | |
| CASPT2 (7e,6o), 0-IPEA | 6-31G(d,p) | 1.381 | 1.438 | n/d |
| CASPT2 (7e,6o), 0-IPEA | cc-pVDZ | 1.391 | 1.448 | n/d |
| CASPT2 (7e,6o), 0-IPEA | cc-pVTZ | 1.380 | 1.436 | n/d |
| CASPT2 (7e,6o), S-IPEA | 6-31G(d,p) | 1.380 | 1.437 | n/d |
| CASPT2 (7e,6o), S-IPEA | cc-pVDZ | 1.390 | 1.447 | n/d |
| CASPT2 (7e,6o), S-IPEA | cc-pVTZ | 1.378 | 1.435 | n/d |
| CASSCF (7e,6o) | 6-31G(d,p) | 1.376 | 1.436 | 2 [ -286, -132 cm$^{-1}$] |
| CASSCF (7e,6o) | cc-pVDZ | 1.379 | 1.438 | 1 [211 cm$^{-1}$] |
| CASSCF (7e,6o) | cc-pVTZ | 1.373 | 1.432 | 1 [218 cm$^{-1}$] |
| CASSCF (7e,6o) | cc-pVQZ | 1.373 | 1.431 | 2 [ -264, -123 cm$^{-1}$] |
| CC2 | cc-pVDZ | 1.393 | 1.446 | 2 [-294, -243 cm$^{-1}$] |
| CC2 | aug-cc-pVDZ | 1.398 | 1.445 | n/d |
| CC2 | cc-pVTZ | 1.381 | 1.435 | 2 [-276, -191 cm$^{-1}$] |
| CC2 | aug-cc-pVTZ | 1.385 | 1.431 | n/d |
| CC2 | cc-pVQZ | 1.380 | 1.432 | 2 [-305, -238 cm$^{-1}$] |
| CC2 | aug-cc-pVQZ | 1.383 | 1.428 | n/d |
| DFT(B3LYP) | 6-31G(d,p) | 1.378 | 1.440 | 1 [-252 cm$^{-1}$] |
| DFT(B3LYP) | cc-pVDZ | 1.381 | 1.442 | 0 |
| DFT(B3LYP) | cc-pVTZ | 1.373 | 1.434 | 1 [-139 cm$^{-1}$] |
| DFT(B3LYP) | cc-pVQZ | 1.374 | 1.434 | 2 [-231, -53 cm$^{-1}$] |
| DFT(M06-2X) | 6-31G(d,p) | 1.373 | 1.435 | 1 [-119 cm$^{-1}$] |
| DFT(M06-2X) | cc-pVDZ | 1.377 | 1.437 | 0 |
| DFT(M06-2X) | cc-pVTZ | 1.370 | 1.431 | 0 |
| DFT(M06-2X) | cc-pVQZ | 1.370 | 1.430 | 1 [-54 cm$^{-1}$] |
| DFT(CAM-B3LYP) | 6-31G(d,p) | 1.371 | 1.434 | 1 [-125 cm$^{-1}$] |
| DFT(CAM-B3LYP) | cc-pVDZ | 1.374 | 1.437 | 1 [-94 cm$^{-1}$] |
| DFT(CAM-B3LYP) | cc-pVTZ | 1.366 | 1.429 | 0 |
| DFT(CAM-B3LYP) | cc-pVQZ | 1.367 | 1.428 | 1 [-131 cm$^{-1}$] |

n/d - *not determined (because of computational cost or some numerical troubles)*

**Table S2.** Selected bond lengths [Å] between carbon atoms in anti-quinoid-like anionic variant of benzene ring obtained from chosen approaches of quantum chemistry computational methods. C-C bond are labelled by designations introduced by **Figure 1**. If it is indicated, vibrational analysis was carried out for the equilibrium structure. All obtained structures are flat, possible imaginary frequencies relate to the swing of hydrogen atoms off the plane of symmetry, or are numerical artefacts of DFT methodology.

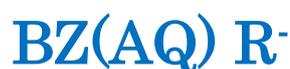

| Total charge: | -1 |
| Spin multiplicity: | 2 |
| Geometry optimization state: | $D_0$ |

| Method | Basis set | Bond lenghts [Å] | | Imag. freq. |
|---|---|---|---|---|
| | | $a^-$ | $b^-$ | |
| CASPT2 (7e,6o), 0-IPEA | 6-31G(d,p) | 1.460 | 1.399 | n/d |
| CASPT2 (7e,6o), 0-IPEA | cc-pVDZ | 1.470 | 1.409 | n/d |
| CASPT2 (7e,6o), 0-IPEA | cc-pVTZ | 1.457 | 1.396 | n/d |
| CASPT2 (7e,6o), S-IPEA | 6-31G(d,p) | 1.459 | 1.398 | n/d |
| CASPT2 (7e,6o), S-IPEA | cc-pVDZ | 1.468 | 1.407 | n/d |
| CASPT2 (7e,6o), S-IPEA | cc-pVTZ | 1.456 | 1.395 | n/d |
| CASSCF (7e,6o) | 6-31G(d,p) | 1.457 | 1.394 | 1 [-182 $cm^{-1}$] |
| CASSCF (7e,6o) | cc-pVDZ | 1.459 | 1.399 | 1 [-155 $cm^{-1}$] |
| CASSCF (7e,6o) | cc-pVTZ | 1.453 | 1.392 | 1 [-188 $cm^{-1}$] |
| CASSCF (7e,6o) | cc-pVQZ | 1.452 | 1.392 | 1 [-254 $cm^{-1}$] |
| CC2 | cc-pVDZ | 1.466 | 1.408 | 1 [-535 $cm^{-1}$] |
| CC2 | aug-cc-pVDZ | 1.461 | 1.412 | n/d |
| CC2 | cc-pVTZ | 1.454 | 1.397 | 1 [-558 $cm^{-1}$] |
| CC2 | aug-cc-pVTZ | 1.447 | 1.399 | n/d |
| CC2 | cc-pVQZ | 1.451 | 1.395 | 1 [-563 $cm^{-1}$] |
| CC2 | aug-cc-pVQZ | 1.443 | 1.396 | n/d |
| DFT(B3LYP) | 6-31G(d,p) | 1.463 | 1.397 | 1 [-220 $cm^{-1}$] |
| DFT(B3LYP) | cc-pVDZ | 1.465 | 1.400 | 1 [-215 $cm^{-1}$] |
| DFT(B3LYP) | cc-pVTZ | 1.457 | 1.392 | 1 [-168 $cm^{-1}$] |
| DFT(B3LYP) | cc-pVQZ | 1.456 | 1.393 | 1 [-168 $cm^{-1}$] |
| DFT(M06-2X) | 6-31G(d,p) | 1.458 | 1.392 | 1 [-322 $cm^{-1}$] |
| DFT(M06-2X) | cc-pVDZ | 1.459 | 1.395 | 1 [-316 $cm^{-1}$] |
| DFT(M06-2X) | cc-pVTZ | 1.454 | 1.389 | 1 [-295 $cm^{-1}$] |
| DFT(M06-2X) | cc-pVQZ | 1.452 | 1.389 | 1 [-355 $cm^{-1}$] |
| DFT(CAM-B3LYP) | 6-31G(d,p) | 1.457 | 1.391 | 1 [-340 $cm^{-1}$] |
| DFT(CAM-B3LYP) | cc-pVDZ | 1.460 | 1.394 | 0 |
| DFT(CAM-B3LYP) | cc-pVTZ | 1.452 | 1.386 | 1 [-314 $cm^{-1}$] |
| DFT(CAM-B3LYP) | cc-pVQZ | 1.451 | 1.386 | 1 [-322 $cm^{-1}$] |

*n/d - not determined (because of computational cost or some numerical troubles)*

**Table S3.** Selected bond lengths [Å] between carbon atoms in quinoid-like cationic variant of benzene ring obtained from chosen approaches of quantum chemistry computational methods. C-C bond are labelled by designations introduced by **Figure 1**. If it is indicated, vibrational analysis was carried out for the equilibrium structure. All obtained structures are flat, possible imaginary frequencies relate to the swing of hydrogen atoms off the plane of symmetry, or are numerical artefacts of DFT methodology.

**Total charge:** 1
**Spin multiplicity:** 2
**Geometry optimization state:** $D_0$

**BZ(Q) R+**

| Method | Basis set | Bond lengths [Å] $a^+$ | $b^+$ | Imag. freq. |
|---|---|---|---|---|
| CASPT2 (5e,6o), 0-IPEA | 6-31G(d,p) | 1.376 | 1.430 | n/d |
| CASPT2 (5e,6o), 0-IPEA | cc-pVDZ | 1.385 | 1.438 | n/d |
| CASPT2 (5e,6o), 0-IPEA | cc-pVTZ | 1.372 | 1.426 | n/d |
| CASPT2 (5e,6o), S-IPEA | 6-31G(d,p) | 1.375 | 1.429 | n/d |
| CASPT2 (5e,6o), S-IPEA | cc-pVDZ | 1.384 | 1.438 | n/d |
| CASPT2 (5e,6o), S-IPEA | cc-pVTZ | 1.370 | 1.426 | n/d |
| CASSCF (5e,6o) | 6-31G(d,p) | 1.370 | 1.426 | 0 |
| CASSCF (5e,6o) | cc-pVDZ | 1.372 | 1.427 | 0 |
| CASSCF (5e,6o) | cc-pVTZ | 1.365 | 1.422 | 0 |
| CASSCF (5e,6o) | cc-pVQZ | 1.364 | 1.421 | 0 |
| CC2 | cc-pVDZ | 1.373 | 1.426 | 0 |
| CC2 | aug-cc-pVDZ | 1.373 | 1.428 | 0 |
| CC2 | cc-pVTZ | 1.359 | 1.414 | 0 |
| CC2 | aug-cc-pVTZ | 1.359 | 1.415 | 0 |
| CC2 | cc-pVQZ | 1.356 | 1.412 | 0 |
| CC2 | aug-cc-pVQZ | 1.357 | 1.413 | 0 |
| DFT(B3LYP) | 6-31G(d,p) | 1.372 | 1.432 | 0 |
| DFT(B3LYP) | cc-pVDZ | 1.374 | 1.433 | 0 |
| DFT(B3LYP) | cc-pVTZ | 1.365 | 1.426 | 0 |
| DFT(B3LYP) | cc-pVQZ | 1.364 | 1.426 | 0 |
| DFT(M06-2X) | 6-31G(d,p) | 1.367 | 1.430 | 0 |
| DFT(M06-2X) | cc-pVDZ | 1.368 | 1.431 | 0 |
| DFT(M06-2X) | cc-pVTZ | 1.360 | 1.425 | 0 |
| DFT(M06-2X) | cc-pVQZ | 1.360 | 1.424 | 0 |
| DFT(CAM-B3LYP) | 6-31G(d,p) | 1.365 | 1.427 | 0 |
| DFT(CAM-B3LYP) | cc-pVDZ | 1.367 | 1.428 | 0 |
| DFT(CAM-B3LYP) | cc-pVTZ | 1.358 | 1.421 | 0 |
| DFT(CAM-B3LYP) | cc-pVQZ | 1.357 | 1.420 | 0 |

n/d - *not determined (because of computational cost or some numerical troubles)*

**Table S4.** Selected bond lengths [Å] between carbon atoms in antiquinoid-like cationic variant of benzene ring obtained from chosen approaches of quantum chemistry computational methods. C-C bond are labelled by designations introduced by **Figure 1**. If it is indicated, vibrational analysis was carried out for the equilibrium structure. All obtained structures are flat, possible imaginary frequencies relate to the swing of hydrogen atoms off the plane of symmetry, or are numerical artefacts of DFT methodology.

**BZ(AQ) R+**

Total charge: 1
Spin multiplicity: 2
Geometry optimization state: $D_0$

| Method | Basis set | Bond lenghts [Å] a+ | Bond lenghts [Å] b+ | Imag. freq. |
|---|---|---|---|---|
| CASPT2 (5e,6o), 0-IPEA | 6-31G(d,p) | 1.451 | 1.393 | n/d |
| CASPT2 (5e,6o), 0-IPEA | cc-pVDZ | 1.459 | 1.402 | n/d |
| CASPT2 (5e,6o), 0-IPEA | cc-pVTZ | 1.448 | 1.389 | n/d |
| CASPT2 (5e,6o), S-IPEA | 6-31G(d,p) | 1.450 | 1.392 | n/d |
| CASPT2 (5e,6o), S-IPEA | cc-pVDZ | 1.458 | 1.400 | n/d |
| CASPT2 (5e,6o), S-IPEA | cc-pVTZ | 1.447 | 1.387 | n/d |
| CASSCF (5e,6o) | 6-31G(d,p) | 1.444 | 1.389 | 1 [-436 cm$^{-1}$] |
| CASSCF (5e,6o) | cc-pVDZ | 1.445 | 1.390 | 1 [-417 cm$^{-1}$] |
| CASSCF (5e,6o) | cc-pVTZ | 1.440 | 1.384 | 1 [-378 cm$^{-1}$] |
| CASSCF (5e,6o) | cc-pVQZ | 1.440 | 1.383 | 1 [-406 cm-1] |
| CC2 | cc-pVDZ | 1.455 | 1.386 | 1 [-1049 cm$^{-1}$] |
| CC2 | aug-cc-pVDZ | 1.457 | 1.387 | 1 [-1032 cm$^{-1}$] |
| CC2 | cc-pVTZ | 1.444 | 1.373 | 1 [-1083 cm$^{-1}$] |
| CC2 | aug-cc-pVTZ | 1.444 | 1.373 | 2 [-1074, -522 cm$^{-1}$] |
| CC2 | cc-pVQZ | 1.441 | 1.370 | 1 [-1092 cm$^{-1}$] |
| CC2 | aug-cc-pVQZ | 1.441 | 1.371 | n/d |
| DFT(B3LYP) | 6-31G(d,p) | 1.453 | 1.391 | 1 [-436 cm$^{-1}$] |
| DFT(B3LYP) | cc-pVDZ | 1.455 | 1.392 | 1 [-434 cm$^{-1}$] |
| DFT(B3LYP) | cc-pVTZ | 1.448 | 1.384 | 1 [-410 cm$^{-1}$] |
| DFT(B3LYP) | cc-pVQZ | 1.447 | 1.383 | 1 [-409 cm-1] |
| DFT(M06-2X) | 6-31G(d,p) | 1.451 | 1.386 | 1 [-1213 cm$^{-1}$] |
| DFT(M06-2X) | cc-pVDZ | 1.452 | 1.388 | 1 [-1166 cm$^{-1}$] |
| DFT(M06-2X) | cc-pVTZ | 1.447 | 1.380 | 1 [-1087 cm$^{-1}$] |
| DFT(M06-2X) | cc-pVQZ | 1.446 | 1.380 | 1 [-1114 cm$^{-1}$] |
| DFT(CAM-B3LYP) | 6-31G(d,p) | 1.448 | 1.384 | 1 [-395 cm$^{-1}$] |
| DFT(CAM-B3LYP) | cc-pVDZ | 1.450 | 1.386 | 1 [-387 cm$^{-1}$] |
| DFT(CAM-B3LYP) | cc-pVTZ | 1.443 | 1.377 | 1 [-370 cm$^{-1}$] |
| DFT(CAM-B3LYP) | cc-pVQZ | 1.443 | 1.377 | 1 [-366 cm$^{-1}$] |

n/d - *not determined (because of computational cost or some numerical troubles)*

**Table S5.** Selected bond lengths [Å] between carbon atoms in neutral variant of benzene ring obtained from chosen approaches of quantum chemistry computational methods. C-C bond are labelled by designations introduced by **Figure 1**. If it is indicated, vibrational analysis was carried out for the equilibrium structure. All obtained structures are flat, possible imaginary frequencies relate to the swing of hydrogen atoms off the plane of symmetry, or are numerical artefacts of DFT methodology..

Total charge: 0
Spin multiplicity: 1
Geometry optimization state: $S_0$

BZ $R^0$

| Method | Basis set | Bond lenght [Å] $a^0=b^0$ | Imag. freq. |
|---|---|---|---|
| CASPT2 (6e,6o), 0-IPEA | 6-31G(d,p) | 1.398 | n/d |
| CASPT2 (6e,6o), 0-IPEA | cc-pVDZ | 1.407 | n/d |
| CASPT2 (6e,6o), 0-IPEA | cc-pVTZ | 1.395 | n/d |
| CASPT2 (6e,6o), S-IPEA | 6-31G(d,p) | 1.397 | n/d |
| CASPT2 (6e,6o), S-IPEA | cc-pVDZ | 1.406 | n/d |
| CASPT2 (6e,6o), S-IPEA | cc-pVTZ | 1.393 | n/d |
| CASSCF (6e,6o) | 6-31G(d,p) | 1.396 | 0 |
| CASSCF (6e,6o) | cc-pVDZ | 1.398 | 0 |
| CASSCF (6e,6o) | cc-pVTZ | 1.392 | 0 |
| CASSCF (6e,6o) | cc-pVQZ | 1.392 | 0 |
| CC2 | cc-pVDZ | 1.407 | 0 |
| CC2 | aug-cc-pVDZ | 1.409 | 0 |
| CC2 | cc-pVTZ | 1.395 | 0 |
| CC2 | aug-cc-pVTZ | 1.396 | 1 [-423 cm$^{-1}$] |
| CC2 | cc-pVQZ | 1.394 | 0 |
| CC2 | aug-cc-pVQZ | 1.394 | 0 |
| DFT(B3LYP) | 6-31G(d,p) | 1.396 | 0 |
| DFT(B3cd LYP) | cc-pVDZ | 1.399 | 0 |
| DFT(B3LYP) | cc-pVTZ | 1.391 | 0 |
| DFT(B3LYP) | cc-pVQZ | 1.391 | 0 |
| DFT(M06-2X) | 6-31G(d,p) | 1.393 | 0 |
| DFT(M06-2X) | cc-pVDZ | 1.395 | 0 |
| DFT(M06-2X) | cc-pVTZ | 1.388 | 0 |
| DFT(M06-2X) | cc-pVQZ | 1.388 | 0 |
| DFT(CAM-B3LYP) | 6-31G(d,p) | 1.391 | 0 |
| DFT(CAM-B3LYP) | cc-pVDZ | 1.393 | 0 |
| DFT(CAM-B3LYP) | cc-pVTZ | 1.385 | 0 |
| DFT(CAM-B3LYP) | cc-pVQZ | 1.385 | 0 |

n/d - *not determined (because of computational cost or some numerical troubles)*

**Table S6.** Selected bond lengths [Å] between carbon atoms in quinoid-like neutral variant of benzene ring in the first electronic triplet excided state obtained from chosen approaches of quantum chemistry computational methods. C-C bond are labelled by designations introduced by **Figure 1**. If it is indicated, vibrational analysis was carried out for the equilibrium structure. All obtained structures are flat, possible imaginary frequencies relate to the swing of hydrogen atoms off the plane of symmetry, or are numerical artefacts of DFT methodology.

|  |  |  |  |
|---|---|---|---|
| Total charge: | 0 | | **BZ(Q) R$^0_{T1}$** |
| Spin multiplicity: | 1 | | |
| Geometry optimization state: | T$_1$ | | |

| Method | Basis set | Bond lenghts [Å] a$^0_{T1}$ / b$^0_{T1}$ | Imag. freq. |
|---|---|---|---|
| CASPT2 (6e,6o), 0-IPEA | 6-31G(d,p) | 1.373  1.461 | n/d |
| CASPT2 (6e,6o), 0-IPEA | cc-pVDZ | 1.382  1.469 | n/d |
| CASPT2 (6e,6o), 0-IPEA | cc-pVTZ | 1.366  1.459 | n/d |
| CASPT2 (6e,6o), S-IPEA | 6-31G(d,p) | 1.365  1.464 | n/d |
| CASPT2 (6e,6o), S-IPEA | cc-pVDZ | 1.373  1.473 | n/d |
| CASPT2 (6e,6o), S-IPEA | cc-pVTZ | 1.358  1.463 | n/d |
| CASSCF (6e,6o) | 6-31G(d,p) | 1.361  1.468 | n/d |
| CASSCF (6e,6o) | cc-pVDZ | 1.363  1.470 | n/d |
| CASSCF (6e,6o) | cc-pVTZ | 1.354  1.466 | n/d |
| CASSCF (6e,6o) | cc-pVQZ | 1.353  1.466 | n/d |
| CC2 | cc-pVDZ | 1.374  1.476 | n/d |
| CC2 | aug-cc-pVDZ | 1.375  1.478 | n/d |
| CC2 | cc-pVTZ | 1.358  1.466 | n/d |
| CC2 | aug-cc-pVTZ | 1.358  1.466 | n/d |
| CC2 | cc-pVQZ | 1.355  1.464 | n/d |
| CC2 | aug-cc-pVQZ | 1.356  1.464 | n/d |
| DFT(B3LYP) | 6-31G(d,p) | 1.347  1.476 | 2 [-543, -81 cm$^{-1}$] |
| DFT(B3LYP) | cc-pVDZ | 1.350  1.477 | 2 [-529, -65 cm$^{-1}$] |
| DFT(B3LYP) | cc-pVTZ | 1.341  1.471 | 2 [-462, -85 cm$^{-1}$] |
| DFT(B3LYP) | cc-pVQZ | 1.341  1.470 | 2 [-458, -89 cm$^{-1}$] |
| DFT(M06-2X) | 6-31G(d,p) | 1.341  1.474 | 1 [-98 cm$^{-1}$] |
| DFT(M06-2X) | cc-pVDZ | 1.344  1.475 | 1 [-94 cm$^{-1}$] |
| DFT(M06-2X) | cc-pVTZ | 1.336  1.470 | 1 [-96 cm$^{-1}$] |
| DFT(M06-2X) | cc-pVQZ | 1.335  1.470 | 1 [-90 cm$^{-1}$] |
| DFT(CAM-B3LYP) | 6-31G(d,p) | 1.339  1.472 | 2 [-636, -78 cm$^{-1}$] |
| DFT(CAM-B3LYP) | cc-pVDZ | 1.342  1.474 | 2 [-620, -63 cm$^{-1}$] |
| DFT(CAM-B3LYP) | cc-pVTZ | 1.333  1.467 | 2 [-523, -82 cm$^{-1}$] |
| DFT(CAM-B3LYP) | cc-pVQZ | 1.333  1.467 | 2 [-516, -86 cm$^{-1}$] |

n/d - *not determined (because of computational cost or some numerical troubles)*

**Table S7.** Selected bond lengths [Å] between carbon atoms in antiquinoid-like neutral variant of benzene ring in the first electronic triplet excided state obtained from chosen approaches of quantum chemistry computational methods. C-C bond are labelled by designations introduced by **Figure 1**. If it is indicated, vibrational analysis was carried out for the equilibrium structure. All obtained structures are flat, possible imaginary frequencies relate to the swing of hydrogen atoms off the plane of symmetry, or are numerical artefacts of DFT methodology.

| | | | | |
|---|---|---|---|---|
| Total charge: | | 0 | | |
| Spin multiplicity: | | 1 | | **BZ(AQ) $R^0_{T1}$** |
| Geometry optimization state: | | $T_1$ | | |

| Method | Basis set | Bond lenghts [Å] | | Imag. freq. |
|---|---|---|---|---|
| | | $a^0_{T1}$ | $b^0_{T1}$ | |
| CASPT2 (6e,6o), 0-IPEA | 6-31G(d,p) | 1.487 | 1.403 | n/d |
| CASPT2 (6e,6o), 0-IPEA | cc-pVDZ | 1.496 | 1.412 | n/d |
| CASPT2 (6e,6o), 0-IPEA | cc-pVTZ | 1.491 | 1.397 | n/d |
| CASPT2 (6e,6o), S-IPEA | 6-31G(d,p) | 1.497 | 1.398 | n/d |
| CASPT2 (6e,6o), S-IPEA | cc-pVDZ | 1.507 | 1.406 | n/d |
| CASPT2 (6e,6o), S-IPEA | cc-pVTZ | 1.499 | 1.392 | n/d |
| CASSCF (6e,6o) | 6-31G(d,p) | 1.499 | 1.398 | n/d |
| CASSCF (6e,6o) | cc-pVDZ | 1.501 | 1.400 | n/d |
| CASSCF (6e,6o) | cc-pVTZ | 1.500 | 1.392 | n/d |
| CASSCF (6e,6o) | cc-pVQZ | 1.499 | 1.391 | n/d |
| CC2 | cc-pVDZ | 1.511 | 1.407 | n/d |
| CC2 | aug-cc-pVDZ | 1.513 | 1.408 | n/d |
| CC2 | cc-pVTZ | 1.505 | 1.393 | n/d |
| CC2 | aug-cc-pVTZ | 1.505 | 1.393 | n/d |
| CC2 | cc-pVQZ | 1.503 | 1.390 | n/d |
| CC2 | aug-cc-pVQZ | 1.503 | 1.391 | n/d |
| DFT(B3LYP) | 6-31G(d,p) | 1.522 | 1.389 | 0 |
| DFT(B3LYP) | cc-pVDZ | 1.523 | 1.391 | 0 |
| DFT(B3LYP) | cc-pVTZ | 1.517 | 1.383 | 0 |
| DFT(B3LYP) | cc-pVQZ | 1.517 | 1.382 | 0 |
| DFT(M06-2X) | 6-31G(d,p) | 1.518 | 1.384 | 1 [-1199 cm-1] |
| DFT(M06-2X) | cc-pVDZ | 1.519 | 1.387 | 1 [-1166cm-1] |
| DFT(M06-2X) | cc-pVTZ | 1.515 | 1.380 | 1 [-1060cm-1] |
| DFT(M06-2X) | cc-pVQZ | 1.514 | 1.379 | 1 [-1131cm-1] |
| DFT(CAM-B3LYP) | 6-31G(d,p) | 1.518 | 1.383 | 0 |
| DFT(CAM-B3LYP) | cc-pVDZ | 1.519 | 1.386 | 0 |
| DFT(CAM-B3LYP) | cc-pVTZ | 1.513 | 1.377 | 0 |
| DFT(CAM-B3LYP) | cc-pVQZ | 1.513 | 1.377 | 0 |

n/d - *not determined (because of computational cost or some numerical troubles)*

**Table S8.** Grochala, Albrecht, and Hoffmann Bond Length Rule formula values [Å] for selected carbon-carbon bonds in quinoid-like benzene variants. C-C bonds are labelled by designations introduced by **Figure 1**. The GAH rule is fulfilled more precisely when the value calculated according the formula is closer to zero. The charge of the molecule and its spin multiplicity can affect each bond length. Therefore unsigned relative percentage values are given in brackets. They were calculated by dividing the absolute value between the difference in length of a given bond in its longest and shortest form.

## BZ(Q) ΔGAH(R)

| Method | Basis set | ΔGAH(R) = $R^+ + R^- - R^0 - R^0_{T1}$ [Å] | | | |
|---|---|---|---|---|---|
| | | a | | b | |
| CASPT2, 0-IPEA | 6-31G(d,p) | -0.014 | (55.8%) | 0.010 | (16.0%) |
| CASPT2, 0-IPEA | cc-pVDZ | -0.013 | (51.8%) | 0.010 | (16.2%) |
| CASPT2, 0-IPEA | cc-pVTZ | -0.009 | (31.0%) | 0.008 | (12.5%) |
| CASPT2, S-IPEA | 6-31G(d,p) | -0.007 | (22.2%) | 0.006 | (8.3%) |
| CASPT2, S-IPEA | cc-pVDZ | -0.006 | (17.9%) | 0.006 | (8.2%) |
| CASPT2, 0-IPEA | cc-pVTZ | -0.003 | (8.6%) | 0.005 | (7.1%) |
| CASSCF | 6-31G(d,p) | -0.012 | (32.9%) | -0.002 | (3.4%) |
| CASSCF | cc-pVDZ | -0.011 | (29.7%) | -0.003 | (3.8%) |
| CASSCF | cc-pVTZ | -0.008 | (22.1%) | -0.004 | (5.2%) |
| CASSCF | cc-pVQZ | -0.008 | (20.3%) | -0.005 | (6.5%) |
| CC2 | cc-pVDZ | -0.016 | (47.3%) | -0.011 | (15.6%) |
| CC2 | aug-cc-pVDZ | -0.013 | (36.7%) | -0.014 | (20.9%) |
| CC2 | cc-pVTZ | -0.013 | (35.9%) | -0.013 | (18.4%) |
| CC2 | aug-cc-pVTZ | -0.011 | (28.6%) | -0.017 | (24.1%) |
| CC2 | cc-pVQZ | -0.013 | (33.7%) | -0.014 | (20.2%) |
| CC2 | aug-cc-pVQZ | -0.010 | (27.3%) | -0.017 | (24.6%) |
| DFT(B3LYP) | 6-31G(d,p) | 0.007 | (13.8%) | -0.001 | (1.1%) |
| DFT(B3LYP) | cc-pVDZ | 0.006 | (13.3%) | -0.001 | (1.4%) |
| DFT(B3LYP) | cc-pVTZ | 0.007 | (13.6%) | -0.001 | (1.8%) |
| DFT(B3LYP) | cc-pVQZ | 0.007 | (14.0%) | -0.002 | (2.1%) |
| DFT(M06-2X) | 6-31G(d,p) | 0.006 | (11.8%) | -0.002 | (2.8%) |
| DFT(M06-2X) | cc-pVDZ | 0.007 | (12.7%) | -0.002 | (2.9%) |
| DFT(M06-2X) | cc-pVTZ | 0.006 | (11.6%) | -0.003 | (3.2%) |
| DFT(M06-2X) | cc-pVQZ | 0.007 | (12.6%) | -0.003 | (3.3%) |
| DFT(CAM-B3LYP) | 6-31G(d,p) | 0.006 | (11.8%) | -0.002 | (3.1%) |
| DFT(CAM-B3LYP) | cc-pVDZ | 0.006 | (11.6%) | -0.001 | (1.6%) |
| DFT(CAM-B3LYP) | cc-pVTZ | 0.006 | (11.7%) | -0.003 | (3.2%) |
| DFT(CAM-B3LYP) | cc-pVQZ | 0.006 | (12.3%) | -0.003 | (3.8%) |

**Table S9.** Grochala, Albrecht, and Hoffmann Bond Length Rule formula values [Å] for selected carbon-carbon bonds in anti-quinoid-like benzene variants. C-C bonds are labelled by designations introduced by **Figure 1**. The GAH rule is fulfilled more precisely when the value calculated according the formula is closer to zero. The charge of the molecule and its spin multiplicity can affect each bond length. Therefore, absolute values of relative percentage values are given in brackets. They were calculated by dividing the absolute value between the difference in length of a given bond in its longest and shortest form. TBD – to be determined.

## BZ(AQ) ΔGAH(R)

| Method | Basis set | $\Delta GAH(R) = R^+ + R^- - R^0 - R^0_{T1}$ [Å] | | | |
|---|---|---|---|---|---|
| | | a | | B | |
| CASPT2, 0-IPEA | 6-31G(d,p) | 0.026 | (29.1%) | -0.010 | (90.9%) |
| CASPT2, 0-IPEA | cc-pVDZ | 0.025 | (28.4%) | -0.009 | (84.5%) |
| CASPT2, 0-IPEA | cc-pVTZ | TBD | TBD | TBD | TBD |
| CASPT2, S-IPEA | 6-31G(d,p) | 0.015 | (14.9%) | -0.005 | (83.9%) |
| CASPT2, S-IPEA | cc-pVDZ | 0.014 | (13.4%) | -0.010 | (100.0%) |
| CASPT2, S-IPEA | cc-pVTZ | TBD | TBD | TBD | TBD |
| CASSCF | 6-31G(d,p) | 0.006 | (5.7%) | -0.011 | (122.8%) |
| CASSCF | cc-pVDZ | 0.005 | (4.4%) | -0.008 | (92.6%) |
| CASSCF | cc-pVTZ | 0.002 | (2.0%) | -0.008 | (101.2%) |
| CASSCF | cc-pVQZ | 0.001 | (1.0%) | -0.008 | (93.1%) |
| CC2 | cc-pVDZ | 0.003 | (2.4%) | -0.020 | (92.9%) |
| CC2 | aug-cc-pVDZ | -0.005 | (4.7%) | -0.019 | (74.4%) |
| CC2 | cc-pVTZ | -0.002 | (2.0%) | -0.019 | (78.3%) |
| CC2 | aug-cc-pVTZ | -0.009 | (8.6%) | -0.018 | (69.3%) |
| CC2 | cc-pVQZ | -0.004 | (4.1%) | -0.018 | (74.3%) |
| CC2 | aug-cc-pVQZ | -0.012 | (11.0%) | -0.017 | (67.7%) |
| DFT(B3LYP) | 6-31G(d,p) | -0.002 | (1.7%) | 0.003 | (31.0%) |
| DFT(B3LYP) | cc-pVDZ | -0.002 | (1.9%) | 0.003 | (29.2%) |
| DFT(B3LYP) | cc-pVTZ | -0.003 | (2.5%) | 0.003 | (28.6%) |
| DFT(B3LYP) | cc-pVQZ | -0.004 | (3.2%) | 0.003 | (28.7%) |
| DFT(M06-2X) | 6-31G(d,p) | -0.002 | (1.4%) | 0.002 | (18.8%) |
| DFT(M06-2X) | cc-pVDZ | -0.002 | (1.7%) | 0.002 | (18.6%) |
| DFT(M06-2X) | cc-pVTZ | -0.002 | (2.0%) | 0.002 | (17.0%) |
| DFT(M06-2X) | cc-pVQZ | -0.003 | (2.5%) | 0.002 | (19.6%) |
| DFT(CAM-B3LYP) | 6-31G(d,p) | -0.003 | (2.4%) | 0.001 | (9.3%) |
| DFT(CAM-B3LYP) | cc-pVDZ | -0.002 | (1.9%) | 0.001 | (16.3%) |
| DFT(CAM-B3LYP) | cc-pVTZ | -0.004 | (3.1%) | 0.001 | (10.1%) |
| DFT(CAM-B3LYP) | cc-pVQZ | -0.005 | (3.5%) | 0.001 | (12.8%) |

**Table S10.** Selected bond lengths [Å] between carbon atoms in anionic variant of cyclobutadiene (CBDE) ring obtained from chosen approaches of quantum chemistry computational methods. C-C bond are labelled by designations introduced by **Figure 1**. If it is indicated, vibrational analysis was carried out for the equilibrium structure. All obtained structures are flat, possible imaginary frequencies relate to the swing of hydrogen atoms off the plane of symmetry, or are numerical artefacts of DFT methodology. All internal angles in carbon ring are right.

**CBDE R$^-$**

Total charge: -1
Spin multiplicity: 2
Geometry optimization state: D$_0$

| Method | Basis set | Bond lengths [Å] a$^-$ | Bond lengths [Å] b$^-$ | Imag. freq. |
|---|---|---|---|---|
| CASPT2 (5e,4o), 0-IPEA | 6-31G(d,p) | 1.510 | 1.401 | n/d |
| CASPT2 (5e,4o), 0-IPEA | cc-pVDZ | 1.524 | 1.413 | n/d |
| CASPT2 (5e,4o), 0-IPEA | cc-pVTZ | 1.514 | 1.401 | n/d |
| CASPT2 (5e,4o), S-IPEA | 6-31G(d,p) | 1.509 | 1.398 | n/d |
| CASPT2 (5e,4o), S-IPEA | cc-pVDZ | 1.522 | 1.418 | n/d |
| CASPT2 (5e,4o), S-IPEA | cc-pVTZ | 1.512 | 1.398 | n/d |
| CASSCF (5e,4o) | 6-31G(d,p) | 1.500 | 1.393 | 3 [-544, -397, -166 cm$^{-1}$] |
| CASSCF (5e,4o) | cc-pVDZ | 1.505 | 1.398 | 3 [-495, -300, -265 cm$^{-1}$] |
| CASSCF (5e,4o) | cc-pVTZ | 1.500 | 1.390 | 2 [-489, -290 cm$^{-1}$] |
| CASSCF (5e,4o) | cc-pVQZ | 1.499 | 1.390 | 3 [-506, -337, -106 cm$^{-1}$] |
| CC2 | cc-pVDZ | 1.521 | 1.413 | n/d |
| CC2 | aug-cc-pVDZ | 1.523 | 1.416 | n/d |
| CC2 | cc-pVTZ | 1.507 | 1.396 | n/d |
| CC2 | aug-cc-pVTZ | 1.506 | 1.396 | n/d |
| CC2 | cc-pVQZ | 1.505 | 1.395 | n/d |
| CC2 | aug-cc-pVQZ | 1.504 | 1.395 | n/d |
| DFT(B3LYP) | 6-31G(d,p) | 1.516 | 1.394 | 2 [-543, -374 cm$^{-1}$] |
| DFT(B3LYP) | cc-pVDZ | 1.520 | 1.399 | 2 [-467, -198 cm$^{-1}$] |
| DFT(B3LYP) | cc-pVTZ | 1.513 | 1.390 | 2 [-477, -191 cm$^{-1}$] |
| DFT(B3LYP) | cc-pVQZ | 1.512 | 1.390 | 2 [-508, -301 cm$^{-1}$] |
| DFT(M06-2X) | 6-31G(d,p) | 1.506 | 1.394 | 2 [-473, -188 cm$^{-1}$] |
| DFT(M06-2X) | cc-pVDZ | 1.510 | 1.392 | 1 [-391 cm$^{-1}$] |
| DFT(M06-2X) | cc-pVTZ | 1.506 | 1.385 | 1 [-369 cm$^{-1}$] |
| DFT(M06-2X) | cc-pVQZ | 1.505 | 1.385 | 1 [-395 cm$^{-1}$] |
| DFT(CAM-B3LYP) | 6-31G(d,p) | 1.508 | 1.379 | 2 [-486, -235 cm$^{-1}$] |
| DFT(CAM-B3LYP) | cc-pVDZ | 1.512 | 1.392 | 1 [-411 cm$^{-1}$] |
| DFT(CAM-B3LYP) | cc-pVTZ | 1.505 | 1.384 | 1 [-421 cm$^{-1}$] |
| DFT(CAM-B3LYP) | cc-pVQZ | 1.504 | 1.384 | 2 [-451, -89 cm$^{-1}$] |

n/d - *not determined (because of computational cost or some numerical troubles)*

**Table S11.** Selected bonds lengths [Å] between carbon atoms in cationic variant of cyclobutadiene (CBDE) ring obtained from chosen approaches of quantum chemistry computational methods C-C bond are labelled by designations introduced by **Figure 1**. If it is indicated, vibrational analysis was carried out for the equilibrium structure. All obtained structures are flat. All internal angles in carbon ring are right.

**CBDE R+**

Total charge: 1
Spin multiplicity: 2
Geometry optimization state: $D_0$

| Method | Basis set | Bond lengths [Å] a+ | Bond lengths [Å] b+ | Imag. freq. |
|---|---|---|---|---|
| CASPT2 (3e,4o), 0-IPEA | 6-31G(d,p) | 1.502 | 1.385 | n/d |
| CASPT2 (3e,4o), 0-IPEA | cc-pVDZ | 1.514 | 1.396 | n/d |
| CASPT2 (3e,4o), 0-IPEA | cc-pVTZ | 1.502 | 1.382 | n/d |
| CASPT2 (3e,4o), S-IPEA | 6-31G(d,p) | 1.500 | 1.383 | n/d |
| CASPT2 (3e,4o), S-IPEA | cc-pVDZ | 1.511 | 1.394 | n/d |
| CASPT2 (3e,4o), S-IPEA | cc-pVTZ | 1.500 | 1.380 | n/d |
| CASSCF (3e,4o) | 6-31G(d,p) | 1.485 | 1.378 | 0 |
| CASSCF (3e,4o) | cc-pVDZ | 1.488 | 1.382 | 0 |
| CASSCF (3e,4o) | cc-pVTZ | 1.483 | 1.375 | 0 |
| CASSCF (3e,4o) | cc-pVQZ | 1.482 | 1.374 | 0 |
| CC2 | cc-pVDZ | 1.510 | 1.398 | n/d |
| CC2 | aug-cc-pVDZ | 1.512 | 1.399 | n/d |
| CC2 | cc-pVTZ | 1.494 | 1.380 | n/d |
| CC2 | aug-cc-pVTZ | 1.494 | 1.380 | n/d |
| CC2 | cc-pVQZ | 1.492 | 1.378 | n/d |
| CC2 | aug-cc-pVQZ | 1.492 | 1.378 | n/d |
| DFT(B3LYP) | 6-31G(d,p) | 1.505 | 1.380 | 0 |
| DFT(B3LYP) | cc-pVDZ | 1.508 | 1.383 | 0 |
| DFT(B3LYP) | cc-pVTZ | 1.501 | 1.373 | 0 |
| DFT(B3LYP) | cc-pVQZ | 1.500 | 1.373 | 0 |
| DFT(M06-2X) | 6-31G(d,p) | 1.497 | 1.375 | 0 |
| DFT(M06-2X) | cc-pVDZ | 1.499 | 1.378 | 0 |
| DFT(M06-2X) | cc-pVTZ | 1.494 | 1.370 | 0 |
| DFT(M06-2X) | cc-pVQZ | 1.494 | 1.369 | 0 |
| DFT(CAM-B3LYP) | 6-31G(d,p) | 1.497 | 1.374 | 0 |
| DFT(CAM-B3LYP) | cc-pVDZ | 1.500 | 1.377 | 0 |
| DFT(CAM-B3LYP) | cc-pVTZ | 1.493 | 1.368 | 0 |
| DFT(CAM-B3LYP) | cc-pVQZ | 1.493 | 1.367 | 0 |

n/d - *not determined (because of computational cost or some numerical troubles)*

**Table S12.** Selected bonds lengths [Å] between carbon atoms in neutral variant of cyclobutadiene (CBDE) ring obtained from chosen approaches of quantum chemistry computational methods. C-C bond are labelled by designations introduced by **Figure 1**. If it is indicated, vibrational analysis was carried out for the equilibrium structure. All obtained structures are flat. All internal angles in carbon ring are right.

**CBDE R$^0$**

Total charge: 0
Spin multiplicity: 1
Geometry optimization state: S$_0$

| Method | Basis set | Bond lenghts [Å] | | Imag. freq. |
|---|---|---|---|---|
| | | a$^0$ | b$^0$ | |
| CASPT2 (4e,4o), 0-IPEA | 6-31G(d,p) | 1.547 | 1.360 | n/d |
| CASPT2 (4e,4o), 0-IPEA | cc-pVDZ | 1.560 | 1.371 | n/d |
| CASPT2 (4e,4o), 0-IPEA | cc-pVTZ | 1.552 | 1.355 | n/d |
| CASPT2 (4e,4o), S-IPEA | 6-31G(d,p) | 1.552 | 1.354 | n/d |
| CASPT2 (4e,4o), S-IPEA | cc-pVDZ | 1.566 | 1.365 | n/d |
| CASPT2 (4e,4o), S-IPEA | cc-pVTZ | 1.557 | 1.350 | n/d |
| CASSCF (4e,4o) | 6-31G(d,p) | 1.545 | 1.353 | 0 |
| CASSCF (4e,4o) | cc-pVDZ | 1.550 | 1.357 | 0 |
| CASSCF (4e,4o) | cc-pVTZ | 1.548 | 1.347 | 0 |
| CASSCF (4e,4o) | cc-pVQZ | 1.547 | 1.346 | 0 |
| CC2 | cc-pVDZ | 1.578 | 1.360 | n/d |
| CC2 | aug-cc-pVDZ | 1.578 | 1.362 | n/d |
| CC2 | cc-pVTZ | 1.563 | 1.342 | n/d |
| CC2 | aug-cc-pVTZ | 1.562 | 1.342 | n/d |
| CC2 | cc-pVQZ | 1.561 | 1.340 | n/d |
| CC2 | aug-cc-pVQZ | 1.560 | 1.340 | n/d |
| DFT(B3LYP) | 6-31G(d,p) | 1.578 | 1.335 | 0 |
| DFT(B3LYP) | cc-pVDZ | 1.581 | 1.339 | 0 |
| DFT(B3LYP) | cc-pVTZ | 1.575 | 1.329 | 0 |
| DFT(B3LYP) | cc-pVQZ | 1.574 | 1.329 | 0 |
| DFT(M06-2X) | 6-31G(d,p) | 1.569 | 1.330 | 0 |
| DFT(M06-2X) | cc-pVDZ | 1.572 | 1.334 | 0 |
| DFT(M06-2X) | cc-pVTZ | 1.567 | 1.325 | 0 |
| DFT(M06-2X) | cc-pVQZ | 1.566 | 1.325 | 0 |
| DFT(CAM-B3LYP) | 6-31G(d,p) | 1.570 | 1.329 | 0 |
| DFT(CAM-B3LYP) | cc-pVDZ | 1.573 | 1.333 | 0 |
| DFT(CAM-B3LYP) | cc-pVTZ | 1.567 | 1.323 | 0 |
| DFT(CAM-B3LYP) | cc-pVQZ | 1.566 | 1.323 | 0 |

n/d - *not determined (because of computational cost or some numerical troubles)*

**Table S13.** Selected bonds lengths [Å] between carbon atoms in neutral variant of cyclobutadiene (CBDE) ring in firs electronic triplet excited state obtained from chosen approaches of quantum chemistry computational methods. C-C bond are labelled by designations introduced by **Figure 1**. If it is indicated, vibrational analysis was carried out for the equilibrium structure. All obtained structures are flat. All internal angles in carbon ring are right.

| | Total charge: | 0 | |
|---|---|---|---|
| | Spin multiplicity: | 1 | **CBDE R$^0_{T1}$** |
| | Geometry optimization state: | T$_1$ | |

| Method | Basis set | Bond lenght [Å] $a^0_{T1} = b^0_{T1}$ | Imag. freq. |
|---|---|---|---|
| CASPT2 (4e,4o), 0-IPEA | 6-31G(d,p) | 1.441 | n/d |
| CASPT2 (4e,4o), 0-IPEA | cc-pVDZ | 1.453 | n/d |
| CASPT2 (4e,4o), 0-IPEA | cc-pVTZ | 1.439 | n/d |
| CASPT2 (4e,4o), S-IPEA | 6-31G(d,p) | 1.440 | n/d |
| CASPT2 (4e,4o), S-IPEA | cc-pVDZ | 1.452 | n/d |
| CASPT2 (4e,4o), S-IPEA | cc-pVTZ | 1.438 | n/d |
| CASSCF (4e,4o) | 6-31G(d,p) | 1.439 | 0 |
| CASSCF (4e,4o) | cc-pVDZ | 1.442 | 0 |
| CASSCF (4e,4o) | cc-pVTZ | 1.436 | 0 |
| CASSCF (4e,4o) | cc-pVQZ | 1.435 | 0 |
| CC2 | cc-pVDZ | 1.453 | n/d |
| CC2 | aug-cc-pVDZ | 1.455 | n/d |
| CC2 | cc-pVTZ | 1.436 | n/d |
| CC2 | aug-cc-pVTZ | 1.437 | n/d |
| CC2 | cc-pVQZ | 1.435 | n/d |
| CC2 | aug-cc-pVQZ | 1.435 | n/d |
| DFT(B3LYP) | 6-31G(d,p) | 1.441 | 0 |
| DFT(B3LYP) | cc-pVDZ | 1.444 | 0 |
| DFT(B3LYP) | cc-pVTZ | 1.436 | 0 |
| DFT(B3LYP) | cc-pVQZ | 1.436 | 0 |
| DFT(M06-2X) | 6-31G(d,p) | 1.433 | 0 |
| DFT(M06-2X) | cc-pVDZ | 1.437 | 0 |
| DFT(M06-2X) | cc-pVTZ | 1.431 | 0 |
| DFT(M06-2X) | cc-pVQZ | 1.430 | 0 |
| DFT(CAM-B3LYP) | 6-31G(d,p) | 1.434 | 0 |
| DFT(CAM-B3LYP) | cc-pVDZ | 1.438 | 0 |
| DFT(CAM-B3LYP) | cc-pVTZ | 1.430 | 0 |
| DFT(CAM-B3LYP) | cc-pVQZ | 1.429 | 0 |

n/d - *not determined (because of computational cost or some numerical troubles)*

**Table S14.** Grochala, Albrecht, and Hoffmann Bond Length Rule formula values [Å] for carbon-carbon bonds in cyclobutadiene variants. C-C bonds are labelled by designations introduced by **Figure 1**. The GAH rule is fulfilled more precisely when the value calculated according the formula is closer to zero. The charge of the molecule and its spin multiplicity can affect each bond length. Therefore unsigned relative percentage values are given in brackets. They were calculated by dividing the absolute value between the difference in length of a given bond in its longest and shortest form.

## CBDE $\Delta$GAH(R)

| Method | Basis set | $\Delta$GAH(R) = $R^+ + R^- - R^0 - R^0_{T1}$ [Å] | | | |
|---|---|---|---|---|---|
| | | a | | b | |
| CASPT2, 0-IPEA | 6-31G(d,p) | 0.024 | (22.6%) | -0.015 | (18.5%) |
| CASPT2, 0-IPEA | cc-pVDZ | 0.025 | (23.4%) | -0.015 | (18.3%) |
| CASPT2, 0-IPEA | cc-pVTZ | 0.025 | (22.1%) | -0.011 | (13.1%) |
| CASPT2, S-IPEA | 6-31G(d,p) | 0.017 | (15.2%) | -0.013 | (15.1%) |
| CASPT2, S-IPEA | cc-pVDZ | 0.015 | (13.2%) | -0.005 | (5.7%) |
| CASPT2, S-IPEA | cc-pVTZ | 0.017 | (14.3%) | -0.010 | (11.4%) |
| CASSCF | 6-31G(d,p) | 0.001 | (0.9%) | -0.021 | (22.4%) |
| CASSCF | cc-pVDZ | 0.001 | (0.9%) | -0.019 | (22.4%) |
| CASSCF | cc-pVTZ | -0.001 | (0.9%) | -0.019 | (20.2%) |
| CASSCF | cc-pVQZ | -0.001 | (0.9%) | -0.017 | (19.1%) |
| CC2 | cc-pVDZ | 0.001 | (0.8%) | -0.001 | (1.4%) |
| CC2 | aug-cc-pVDZ | 0.001 | (1.1%) | -0.002 | (1.9%) |
| CC2 | cc-pVTZ | 0.002 | (1.2%) | -0.002 | (1.7%) |
| CC2 | aug-cc-pVTZ | 0.002 | (1.6%) | -0.003 | (2.8%) |
| CC2 | cc-pVQZ | 0.002 | (1.4%) | -0.001 | (1.4%) |
| CC2 | aug-cc-pVQZ | 0.002 | (1.3%) | -0.002 | (2.4%) |
| DFT(B3LYP) | 6-31G(d,p) | 0.002 | (1.5%) | -0.002 | (1.9%) |
| DFT(B3LYP) | cc-pVDZ | 0.003 | (2.2%) | -0.001 | (1.0%) |
| DFT(B3LYP) | cc-pVTZ | 0.003 | (2.2%) | -0.002 | (1.9%) |
| DFT(B3LYP) | cc-pVQZ | 0.002 | (1.4%) | -0.002 | (1.9%) |
| DFT(M06-2X) | 6-31G(d,p) | 0.001 | (0.7%) | 0.006 | (5.8%) |
| DFT(M06-2X) | cc-pVDZ | 0.000 | (0.0%) | -0.001 | (1.0%) |
| DFT(M06-2X) | cc-pVTZ | 0.002 | (1.5%) | -0.001 | (0.9%) |
| DFT(M06-2X) | cc-pVQZ | 0.003 | (2.2%) | -0.001 | (1.0%) |
| DFT(CAM-B3LYP) | 6-31G(d,p) | 0.001 | (0.7%) | -0.010 | (9.5%) |
| DFT(CAM-B3LYP) | cc-pVDZ | 0.001 | (0.7%) | -0.002 | (1.9%) |
| DFT(CAM-B3LYP) | cc-pVTZ | 0.001 | (0.7%) | -0.001 | (0.9%) |
| DFT(CAM-B3LYP) | cc-pVQZ | 0.002 | (1.5%) | -0.001 | (0.9%) |

**Table S15.** Bonds lengths [Å] between carbon atoms in neutral cyclopropenium radical (CP) obtained from selected methods of computational quantum chemistry. Vibrational analysis was carried out for the equilibrium structure

Total charge: 0
Spin multiplicity: 2
Geometry optimization state: $D_0$

| Method | Basis set | Bond lenghts [Å] | | | Imag. freq. |
|---|---|---|---|---|---|
| | | $C_1$–$C_2$ | $C_2$–$C_3$ | $C_1$–$C_3$ | |
| DFT(B3LYP) | 6-31G(d,p) | 1.465 | 1.465 | 1.315 | 0 |
| DFT(B3LYP) | cc-pVDZ | 1.469 | 1.469 | 1.320 | 0 |
| DFT(B3LYP) | cc-pVTZ | 1.456 | 1.456 | 1.308 | 0 |
| DFT(B3LYP) | cc-pVQZ | 1.455 | 1.455 | 1.307 | 0 |
| DFT(M06-2X) | 6-31G(d,p) | 1.458 | 1.458 | 1.310 | 0 |
| DFT(M06-2X) | cc-pVDZ | 1.464 | 1.464 | 1.315 | 0 |
| DFT(M06-2X) | cc-pVTZ | 1.453 | 1.453 | 1.304 | 0 |
| DFT(M06-2X) | cc-pVQZ | 1.452 | 1.452 | 1.304 | 0 |
| DFT(CAM-B3LYP) | 6-31G(d,p) | 1.460 | 1.460 | 1.310 | 0 |
| DFT(CAM-B3LYP) | cc-pVDZ | 1.463 | 1.463 | 1.314 | 0 |
| DFT(CAM-B3LYP) | cc-pVTZ | 1.450 | 1.450 | 1.302 | 0 |
| DFT(CAM-B3LYP) | cc-pVQZ | 1.448 | 1.448 | 1.301 | 0 |

**Table S16.** Bonds lengths [Å] between carbon atoms in cyclopropenium radical bication ($CP^{2+}$) obtained from selected methods of computational quantum chemistry. Vibrational analysis was carried out for the equilibrium structure.

Total charge: 2
Spin multiplicity: 2
Geometry optimization state: $D_0$

| Method | Basis set | Bond lenghts [Å] | | | Imag. freq. |
|---|---|---|---|---|---|
| | | $C_1$–$C_2$ | $C_2$–$C_3$ | $C_1$–$C_3$ | |
| DFT(B3LYP) | 6-31G(d,p) | 1.343 | 1.343 | 1.600 | 0 |
| DFT(B3LYP) | cc-pVDZ | 1.348 | 1.348 | 1.607 | 0 |
| DFT(B3LYP) | cc-pVTZ | 1.336 | 1.336 | 1.599 | 0 |
| DFT(B3LYP) | cc-pVQZ | 1.335 | 1.335 | 1.597 | 0 |
| DFT(M06-2X) | 6-31G(d,p) | 1.341 | 1.341 | 1.572 | 0 |
| DFT(M06-2X) | cc-pVDZ | 1.347 | 1.347 | 1.607 | 0 |
| DFT(M06-2X) | cc-pVTZ | 1.335 | 1.335 | 1.572 | 0 |
| DFT(M06-2X) | cc-pVQZ | 1.335 | 1.335 | 1.570 | 0 |
| DFT(CAM-B3LYP) | 6-31G(d,p) | 1.338 | 1.338 | 1.579 | 0 |
| DFT(CAM-B3LYP) | cc-pVDZ | 1.343 | 1.343 | 1.588 | 0 |
| DFT(CAM-B3LYP) | cc-pVTZ | 1.331 | 1.331 | 1.579 | 0 |
| DFT(CAM-B3LYP) | cc-pVQZ | 1.330 | 1.330 | 1.577 | 0 |

**Table S17.** Bonds lengths [Å] between carbon atoms in cyclopropenium cation (CP$^+$) in the ground singlet state obtained from selected methods of computational quantum chemistry. Vibrational analysis was carried out for the equilibrium structure.

Total charge: 1
Spin multiplicity: 1
Geometry optimization state: $S_0$

| Method | Basis set | Bond lenghts [Å] | | | Imag. freq. |
|---|---|---|---|---|---|
| | | $C_1$–$C_2$ | $C_2$–$C_3$ | $C_1$–$C_3$ | |
| DFT(B3LYP) | 6-31G(d,p) | 1.366 | 1.366 | 1.366 | 0 |
| DFT(B3LYP) | cc-pVDZ | 1.371 | 1.371 | 1.371 | 0 |
| DFT(B3LYP) | cc-pVTZ | 1.358 | 1.358 | 1.358 | 0 |
| DFT(B3LYP) | cc-pVQZ | 1.357 | 1.357 | 1.357 | 0 |
| DFT(M06-2X) | 6-31G(d,p) | 1.362 | 1.362 | 1.362 | 0 |
| DFT(M06-2X) | cc-pVDZ | 1.367 | 1.367 | 1.367 | 0 |
| DFT(M06-2X) | cc-pVTZ | 1.356 | 1.356 | 1.356 | 0 |
| DFT(M06-2X) | cc-pVQZ | 1.355 | 1.355 | 1.355 | 0 |
| DFT(CAM-B3LYP) | 6-31G(d,p) | 1.361 | 1.361 | 1.361 | 0 |
| DFT(CAM-B3LYP) | cc-pVDZ | 1.366 | 1.366 | 1.366 | 0 |
| DFT(CAM-B3LYP) | cc-pVTZ | 1.352 | 1.352 | 1.352 | 0 |
| DFT(CAM-B3LYP) | cc-pVQZ | 1.351 | 1.351 | 1.351 | 0 |

**Table S18.** Bonds lengths [Å] between carbon atoms in cyclopropenium cation (CP$^+$) in the first excited triplet state obtained from selected methods of computational quantum chemistry. Vibrational analysis was carried out for the equilibrium structure.

Total charge: 1
Spin multiplicity: 3
Geometry optimization state: $T_1$

| Method | Basis set | Bond lenghts [Å] | | | Imag. freq. |
|---|---|---|---|---|---|
| | | $C_1$–$C_2$ | $C_2$–$C_3$ | $C_1$–$C_3$ | |
| DFT(B3LYP) | 6-31G(d,p) | 1.322 | 1.370 | 1.956 | 0 |
| DFT(B3LYP) | cc-pVDZ | 1.327 | 1.368 | 1.968 | 0 |
| DFT(B3LYP) | cc-pVTZ | 1.304 | 1.359 | 1.958 | 0 |
| DFT(B3LYP) | cc-pVQZ | 1.303 | 1.359 | 1.955 | 0 |
| DFT(M06-2X) | 6-31G(d,p) | 1.298 | 1.405 | 1.874 | 0 |
| DFT(M06-2X) | cc-pVDZ | 1.302 | 1.405 | 1.880 | 0 |
| DFT(M06-2X) | cc-pVTZ | 1.291 | 1.397 | 1.872 | 0 |
| DFT(M06-2X) | cc-pVQZ | 1.289 | 1.399 | 1.868 | 0 |
| DFT(CAM-B3LYP) | 6-31G(d,p) | 1.310 | 1.374 | 1.892 | 0 |
| DFT(CAM-B3LYP) | cc-pVDZ | 1.316 | 1.376 | 1.903 | 0 |
| DFT(CAM-B3LYP) | cc-pVTZ | 1.304 | 1.365 | 1.894 | 0 |
| DFT(CAM-B3LYP) | cc-pVQZ | 1.303 | 1.365 | 1.890 | 0 |

**Table S19.** Mulliken atomic spin densities obtained with chosen computational approaches for quinoid isomer of anionic form of benzene molecule. In the right part of the table atomic spin densities of hydrogens are summed into heavy atoms they are connected to. Please note, that data for CASPT2 method are computed for equilibrium geometries obtained at this level of theory, however spin densities are computed basing on CASSCF wavefunction.

# BZ(Q) R⁻

| Method | Basis set | H1 | C2 | C3 | H4 | C5 | H6 | C7 | H8 | C9 | H10 | C11 | H12 | H1+C2 | C3+H4 | C5+H6 | C7+H8 | C9+J10 | C11+H12 |
|---|---|---|---|---|---|---|---|---|---|---|---|---|---|---|---|---|---|---|---|
| CASPT2, 0-IPEA | 6-31G(d,p) | 0.0021 | 0.4269 | 0.0352 | 0.0003 | 0.0352 | 0.0003 | 0.4269 | 0.0021 | 0.0352 | 0.0003 | 0.0352 | 0.0003 | 0.4290 | 0.0355 | 0.0355 | 0.4290 | 0.0355 | 0.0355 |
| CASPT2, 0-IPEA | cc-pVDZ | 0.0040 | 0.4205 | 0.0372 | 0.0005 | 0.0372 | 0.0005 | 0.4205 | 0.0040 | 0.0372 | 0.0005 | 0.0372 | 0.0005 | 0.4245 | 0.0377 | 0.0377 | 0.4245 | 0.0377 | 0.0377 |
| CASPT2, 0-IPEA | cc-pVTZ | 0.0089 | 0.4074 | 0.0406 | 0.0012 | 0.0406 | 0.0012 | 0.4074 | 0.0089 | 0.0406 | 0.0012 | 0.0406 | 0.0012 | 0.4163 | 0.0418 | 0.0418 | 0.4163 | 0.0418 | 0.0418 |
| CASPT2, S-IPEA | 6-31G(d,p) | 0.0021 | 0.4268 | 0.0353 | 0.0003 | 0.0353 | 0.0003 | 0.4268 | 0.0021 | 0.0353 | 0.0003 | 0.0353 | 0.0003 | 0.4289 | 0.0356 | 0.0356 | 0.4289 | 0.0356 | 0.0356 |
| CASPT2, S-IPEA | cc-pVDZ | 0.0040 | 0.4204 | 0.0373 | 0.0005 | 0.0373 | 0.0005 | 0.4204 | 0.0040 | 0.0373 | 0.0005 | 0.0373 | 0.0005 | 0.4244 | 0.0378 | 0.0378 | 0.4244 | 0.0378 | 0.0378 |
| CASPT2, S-IPEA | cc-pVTZ | 0.0089 | 0.4073 | 0.0407 | 0.0012 | 0.0407 | 0.0012 | 0.4073 | 0.0089 | 0.0407 | 0.0012 | 0.0407 | 0.0012 | 0.4162 | 0.0419 | 0.0419 | 0.4162 | 0.0419 | 0.0419 |
| CASSCF | 6-31G(d,p) | 0.0021 | 0.4269 | 0.0352 | 0.0003 | 0.0352 | 0.0003 | 0.4269 | 0.0021 | 0.0352 | 0.0003 | 0.0352 | 0.0003 | 0.4290 | 0.0355 | 0.0355 | 0.4290 | 0.0355 | 0.0355 |
| CASSCF | cc-pVDZ | 0.0041 | 0.4194 | 0.0377 | 0.0005 | 0.0377 | 0.0005 | 0.4194 | 0.0041 | 0.0377 | 0.0005 | 0.0377 | 0.0005 | 0.4235 | 0.0382 | 0.0382 | 0.4235 | 0.0382 | 0.0382 |
| CASSCF | cc-pVTZ | 0.0090 | 0.4072 | 0.0407 | 0.0012 | 0.0407 | 0.0012 | 0.4072 | 0.0090 | 0.0407 | 0.0012 | 0.0407 | 0.0012 | 0.4162 | 0.0419 | 0.0419 | 0.4162 | 0.0419 | 0.0419 |
| CASSCF | cc-pVQZ | 0.0157 | 0.4029 | 0.0383 | 0.0024 | 0.0383 | 0.0024 | 0.4029 | 0.0157 | 0.0383 | 0.0024 | 0.0383 | 0.0024 | 0.4186 | 0.0407 | 0.0407 | 0.4186 | 0.0407 | 0.0407 |
| DFT(B3LYP) | 6-31G(d,p) | 0.0021 | 0.3585 | 0.0693 | 0.0004 | 0.0693 | 0.0004 | 0.3585 | 0.0021 | 0.0693 | 0.0004 | 0.0693 | 0.0004 | 0.3606 | 0.0697 | 0.0697 | 0.3606 | 0.0697 | 0.0697 |
| DFT(B3LYP) | cc-pVDZ | 0.0041 | 0.3517 | 0.0712 | 0.0009 | 0.0712 | 0.0009 | 0.3517 | 0.0041 | 0.0712 | 0.0009 | 0.0712 | 0.0009 | 0.3558 | 0.0721 | 0.0721 | 0.3558 | 0.0721 | 0.0721 |
| DFT(B3LYP) | cc-pVTZ | 0.0097 | 0.3414 | 0.0723 | 0.0021 | 0.0723 | 0.0021 | 0.3414 | 0.0097 | 0.0723 | 0.0021 | 0.0723 | 0.0021 | 0.3511 | 0.0745 | 0.0745 | 0.3511 | 0.0745 | 0.0745 |
| DFT(B3LYP) | cc-pVQZ | 0.0188 | 0.3368 | 0.0681 | 0.0041 | 0.0681 | 0.0041 | 0.3368 | 0.0188 | 0.0681 | 0.0041 | 0.0681 | 0.0041 | 0.3556 | 0.0722 | 0.0722 | 0.3556 | 0.0722 | 0.0722 |
| DFT(M06-2X) | 6-31G(d,p) | 0.0018 | 0.3554 | 0.0710 | 0.0004 | 0.0710 | 0.0004 | 0.3554 | 0.0018 | 0.0710 | 0.0004 | 0.0710 | 0.0004 | 0.3572 | 0.0714 | 0.0714 | 0.3572 | 0.0714 | 0.0714 |
| DFT(M06-2X) | cc-pVDZ | 0.0037 | 0.3484 | 0.0731 | 0.0008 | 0.0731 | 0.0008 | 0.3484 | 0.0037 | 0.0731 | 0.0008 | 0.0731 | 0.0008 | 0.3522 | 0.0739 | 0.0739 | 0.3522 | 0.0739 | 0.0739 |
| DFT(M06-2X) | cc-pVTZ | 0.0084 | 0.3389 | 0.0745 | 0.0018 | 0.0745 | 0.0018 | 0.3389 | 0.0084 | 0.0745 | 0.0018 | 0.0745 | 0.0018 | 0.3473 | 0.0763 | 0.0763 | 0.3473 | 0.0763 | 0.0763 |
| DFT(M06-2X) | cc-pVQZ | 0.0141 | 0.3352 | 0.0722 | 0.0032 | 0.0722 | 0.0032 | 0.3352 | 0.0141 | 0.0722 | 0.0032 | 0.0722 | 0.0032 | 0.3493 | 0.0754 | 0.0754 | 0.3493 | 0.0754 | 0.0754 |
| DFT(CAM-B3LYP) | 6-31G(d,p) | 0.0020 | 0.3575 | 0.0698 | 0.0004 | 0.0698 | 0.0004 | 0.3575 | 0.0020 | 0.0698 | 0.0004 | 0.0698 | 0.0004 | 0.3595 | 0.0702 | 0.0702 | 0.3595 | 0.0702 | 0.0702 |
| DFT(CAM-B3LYP) | cc-pVDZ | -0.0261 | 0.4863 | 0.0232 | -0.0033 | 0.0232 | -0.0033 | 0.4863 | -0.0261 | 0.0232 | -0.0033 | 0.0232 | -0.0033 | 0.4602 | 0.0199 | 0.0199 | 0.4602 | 0.0199 | 0.0199 |
| DFT(CAM-B3LYP) | cc-pVTZ | 0.0091 | 0.3402 | 0.0734 | 0.0020 | 0.0734 | 0.0020 | 0.3402 | 0.0091 | 0.0734 | 0.0020 | 0.0734 | 0.0020 | 0.3493 | 0.0754 | 0.0754 | 0.3493 | 0.0754 | 0.0754 |
| DFT(CAM-B3LYP) | cc-pVQZ | 0.0174 | 0.3361 | 0.0694 | 0.0038 | 0.0694 | 0.0038 | 0.3361 | 0.0174 | 0.0694 | 0.0038 | 0.0694 | 0.0038 | 0.3535 | 0.0732 | 0.0732 | 0.3535 | 0.0732 | 0.0732 |

**Table S20.** Mulliken atomic spin densities obtained with chosen computational approaches for antiquinoid isomer of anionic form of benzene molecule. In the right part of the table atomic spin densities of hydrogens are summed into heavy atoms they are connected to. Please note, that data for CASPT2 method are computed for equilibrium geometries obtained at this level of theory, however spin densities are computed basing on CASSCF wavefunction. TBD – to be determined.

BZ(AQ) R⁻

| Method | Basis set | H1 | C2 | C3 | H4 | C5 | H6 | C7 | H8 | C9 | H10 | C11 | H12 | H1+C2 | C3+H4 | C5+H6 | C7+H8 | C9+J10 | C11+H12 |
|---|---|---|---|---|---|---|---|---|---|---|---|---|---|---|---|---|---|---|---|
| CASPT2, 0-IPEA | 6-31G(d,p) | 0.0014 | 0.2931 | -0.0888 | -0.0002 | 0.2931 | 0.0014 | 0.2931 | 0.0014 | -0.0888 | -0.0002 | 0.2931 | 0.0014 | 0.2945 | -0.0890 | 0.2945 | 0.2945 | -0.0890 | 0.2945 |
| CASPT2, 0-IPEA | cc-pVDZ | 0.0027 | 0.2894 | -0.0838 | -0.0004 | 0.2894 | 0.0027 | 0.2894 | 0.0027 | -0.0838 | -0.0004 | 0.2894 | 0.0027 | 0.2921 | -0.0842 | 0.2921 | 0.2921 | -0.0842 | 0.2921 |
| CASPT2, 0-IPEA | cc-pVTZ | | | | | | | | | TBD | | | | | | | | | |
| CASPT2, S-IPEA | 6-31G(d,p) | 0.0014 | 0.2930 | -0.0886 | -0.0002 | 0.2930 | 0.0014 | 0.2930 | 0.0014 | -0.0886 | -0.0002 | 0.2930 | 0.0014 | 0.2944 | -0.0888 | 0.2944 | 0.2944 | -0.0888 | 0.2944 |
| CASPT2, S-IPEA | cc-pVDZ | 0.0027 | 0.2893 | -0.0836 | -0.0004 | 0.2893 | 0.0027 | 0.2893 | 0.0027 | -0.0836 | -0.0004 | 0.2893 | 0.0027 | 0.2920 | -0.0840 | 0.2920 | 0.2920 | -0.0840 | 0.2920 |
| CASPT2, S-IPEA | cc-pVTZ | | | | | | | | | TBD | | | | | | | | | |
| CASSCF | 6-31G(d,p) | 0.0014 | 0.2928 | -0.0883 | -0.0002 | 0.2928 | 0.0014 | 0.2928 | 0.0014 | -0.0883 | -0.0002 | 0.2928 | 0.0014 | 0.2942 | -0.0885 | 0.2942 | 0.2942 | -0.0885 | 0.2942 |
| CASSCF | cc-pVDZ | 0.0028 | 0.2885 | -0.0821 | -0.0004 | 0.2885 | 0.0028 | 0.2885 | 0.0028 | -0.0821 | -0.0004 | 0.2885 | 0.0028 | 0.2913 | -0.0825 | 0.2913 | 0.2913 | -0.0825 | 0.2913 |
| CASSCF | cc-pVTZ | 0.0060 | 0.2804 | -0.0722 | -0.0008 | 0.2804 | 0.0060 | 0.2804 | 0.0060 | -0.0722 | -0.0008 | 0.2804 | 0.0060 | 0.2864 | -0.0730 | 0.2864 | 0.2864 | -0.0730 | 0.2864 |
| CASSCF | cc-pVQZ | 0.0104 | 0.2769 | -0.0738 | -0.0007 | 0.2769 | 0.0104 | 0.2769 | 0.0104 | -0.0738 | -0.0007 | 0.2769 | 0.0104 | 0.2873 | -0.0745 | 0.2873 | 0.2873 | -0.0745 | 0.2873 |
| DFT(B3LYP) | 6-31G(d,p) | 0.0015 | 0.2455 | 0.0061 | 0.0000 | 0.2455 | 0.0015 | 0.2455 | 0.0015 | 0.0061 | 0.0000 | 0.2455 | 0.0015 | 0.2470 | 0.0061 | 0.2470 | 0.2470 | 0.0061 | 0.2470 |
| DFT(B3LYP) | cc-pVDZ | 0.0029 | 0.2418 | 0.0105 | 0.0000 | 0.2418 | 0.0029 | 0.2418 | 0.0029 | 0.0105 | 0.0000 | 0.2418 | 0.0029 | 0.2448 | 0.0105 | 0.2448 | 0.2448 | 0.0105 | 0.2448 |
| DFT(B3LYP) | cc-pVTZ | 0.0070 | 0.2359 | 0.0142 | 0.0000 | 0.2359 | 0.0070 | 0.2359 | 0.0070 | 0.0142 | 0.0000 | 0.2359 | 0.0070 | 0.2429 | 0.0143 | 0.2429 | 0.2429 | 0.0143 | 0.2429 |
| DFT(B3LYP) | cc-pVQZ | 0.0135 | 0.2331 | 0.0067 | 0.0002 | 0.2331 | 0.0135 | 0.2331 | 0.0135 | 0.0067 | 0.0002 | 0.2331 | 0.0135 | 0.2465 | 0.0069 | 0.2465 | 0.2465 | 0.0069 | 0.2465 |
| DFT(M06-2X) | 6-31G(d,p) | 0.0013 | 0.2454 | 0.0065 | 0.0000 | 0.2454 | 0.0013 | 0.2454 | 0.0013 | 0.0065 | 0.0000 | 0.2454 | 0.0013 | 0.2467 | 0.0065 | 0.2467 | 0.2467 | 0.0065 | 0.2467 |
| DFT(M06-2X) | cc-pVDZ | 0.0026 | 0.2416 | 0.0115 | 0.0000 | 0.2416 | 0.0026 | 0.2416 | 0.0026 | 0.0115 | 0.0000 | 0.2416 | 0.0026 | 0.2442 | 0.0115 | 0.2442 | 0.2442 | 0.0115 | 0.2442 |
| DFT(M06-2X) | cc-pVTZ | 0.0060 | 0.2357 | 0.0166 | 0.0000 | 0.2357 | 0.0060 | 0.2357 | 0.0060 | 0.0166 | 0.0000 | 0.2357 | 0.0060 | 0.2417 | 0.0167 | 0.2417 | 0.2417 | 0.0167 | 0.2417 |
| DFT(M06-2X) | cc-pVQZ | 0.0102 | 0.2326 | 0.0143 | 0.0001 | 0.2326 | 0.0102 | 0.2326 | 0.0102 | 0.0143 | 0.0001 | 0.2326 | 0.0102 | 0.2428 | 0.0145 | 0.2428 | 0.2428 | 0.0145 | 0.2428 |
| DFT(CAM-B3LYP) | 6-31G(d,p) | 0.0014 | 0.2453 | 0.0065 | 0.0000 | 0.2453 | 0.0014 | 0.2453 | 0.0014 | 0.0065 | 0.0000 | 0.2453 | 0.0014 | 0.2467 | 0.0065 | 0.2467 | 0.2467 | 0.0065 | 0.2467 |
| DFT(CAM-B3LYP) | cc-pVDZ | -0.0185 | 0.3304 | -0.1281 | 0.0044 | 0.3304 | -0.0185 | 0.3304 | -0.0185 | -0.1281 | 0.0044 | 0.3304 | -0.0185 | 0.3118 | -0.1236 | 0.3118 | 0.3118 | -0.1236 | 0.3118 |
| DFT(CAM-B3LYP) | cc-pVTZ | 0.0065 | 0.2355 | 0.0160 | 0.0000 | 0.2355 | 0.0065 | 0.2355 | 0.0065 | 0.0160 | 0.0000 | 0.2355 | 0.0065 | 0.2420 | 0.0161 | 0.2420 | 0.2420 | 0.0161 | 0.2420 |
| DFT(CAM-B3LYP) | cc-pVQZ | 0.0124 | 0.2325 | 0.0100 | 0.0002 | 0.2325 | 0.0124 | 0.2325 | 0.0124 | 0.0100 | 0.0002 | 0.2325 | 0.0124 | 0.2449 | 0.0102 | 0.2449 | 0.2449 | 0.0102 | 0.2449 |

Table S21. Mulliken atomic spin densities obtained with chosen computational approaches for quinoid isomer of cationic form of benzene molecule. In the right part of the table atomic spin densities of hydrogens are summed into heavy atoms they are connected to. Please note, that data for CASPT2 method are computed for equilibrium geometries obtained at this level of theory, however spin densities are computed basing on CASSCF wavefunction.

BZ(Q) R+

| Method | Basis set | H1 | C2 | C3 | H4 | C5 | H6 | C7 | H8 | C9 | H10 | C11 | H12 | H1+C2 | C3+H4 | C5+H6 | C7+H8 | C9+J10 | C11+H12 |
|---|---|---|---|---|---|---|---|---|---|---|---|---|---|---|---|---|---|---|---|
| CASPT2, 0-IPEA | 6-31G(d,p) | 0.0009 | 0.4341 | 0.0318 | 0.0001 | 0.0330 | 0.0001 | 0.4341 | 0.0009 | 0.0318 | 0.0001 | 0.0330 | 0.0001 | 0.4350 | 0.0319 | 0.0331 | 0.4350 | 0.0319 | 0.0331 |
| CASPT2, 0-IPEA | cc-pVDZ | 0.0021 | 0.4289 | 0.0343 | 0.0002 | 0.0343 | 0.0002 | 0.4289 | 0.0021 | 0.0343 | 0.0002 | 0.0343 | 0.0002 | 0.4310 | 0.0345 | 0.0345 | 0.4310 | 0.0345 | 0.0345 |
| CASPT2, 0-IPEA | cc-pVTZ | 0.0044 | 0.4155 | 0.0396 | 0.0004 | 0.0396 | 0.0004 | 0.4155 | 0.0044 | 0.0396 | 0.0004 | 0.0396 | 0.0004 | 0.4199 | 0.0400 | 0.0400 | 0.4199 | 0.0400 | 0.0400 |
| CASPT2, S-IPEA | 6-31G(d,p) | 0.0009 | 0.4342 | 0.0323 | 0.0001 | 0.0325 | 0.0001 | 0.4342 | 0.0009 | 0.0323 | 0.0001 | 0.0325 | 0.0001 | 0.4351 | 0.0324 | 0.0326 | 0.4351 | 0.0324 | 0.0326 |
| CASPT2, S-IPEA | cc-pVDZ | 0.0021 | 0.4289 | 0.0343 | 0.0002 | 0.0343 | 0.0002 | 0.4289 | 0.0021 | 0.0343 | 0.0002 | 0.0343 | 0.0002 | 0.4310 | 0.0345 | 0.0345 | 0.4310 | 0.0345 | 0.0345 |
| CASPT2, S-IPEA | cc-pVTZ | 0.0044 | 0.4155 | 0.0396 | 0.0004 | 0.0396 | 0.0004 | 0.4155 | 0.0044 | 0.0396 | 0.0004 | 0.0396 | 0.0004 | 0.4199 | 0.0400 | 0.0400 | 0.4199 | 0.0400 | 0.0400 |
| CASSCF | 6-31G(d,p) | 0.0010 | 0.4340 | 0.0324 | 0.0001 | 0.0324 | 0.0001 | 0.4340 | 0.0010 | 0.0324 | 0.0001 | 0.0324 | 0.0001 | 0.4350 | 0.0325 | 0.0325 | 0.4350 | 0.0325 | 0.0325 |
| CASSCF | cc-pVDZ | 0.0022 | 0.4280 | 0.0347 | 0.0002 | 0.0347 | 0.0002 | 0.4280 | 0.0022 | 0.0347 | 0.0002 | 0.0347 | 0.0002 | 0.4302 | 0.0349 | 0.0349 | 0.4302 | 0.0349 | 0.0349 |
| CASSCF | cc-pVTZ | 0.0045 | 0.4152 | 0.0397 | 0.0004 | 0.0397 | 0.0004 | 0.4152 | 0.0045 | 0.0397 | 0.0004 | 0.0397 | 0.0004 | 0.4197 | 0.0401 | 0.0401 | 0.4197 | 0.0401 | 0.0401 |
| CASSCF | cc-pVQZ | 0.0055 | 0.4106 | 0.0414 | 0.0006 | 0.0414 | 0.0006 | 0.4106 | 0.0055 | 0.0414 | 0.0006 | 0.0414 | 0.0006 | 0.4161 | 0.0420 | 0.0420 | 0.4161 | 0.0420 | 0.0420 |
| DFT(B3LYP) | 6-31G(d,p) | 0.0010 | 0.3611 | 0.0688 | 0.0002 | 0.0688 | 0.0002 | 0.3611 | 0.0010 | 0.0688 | 0.0002 | 0.0688 | 0.0002 | 0.3621 | 0.0690 | 0.0690 | 0.3621 | 0.0690 | 0.0690 |
| DFT(B3LYP) | cc-pVDZ | 0.0022 | 0.3565 | 0.0702 | 0.0004 | 0.0702 | 0.0004 | 0.3565 | 0.0022 | 0.0702 | 0.0004 | 0.0702 | 0.0004 | 0.3587 | 0.0707 | 0.0707 | 0.3587 | 0.0707 | 0.0707 |
| DFT(B3LYP) | cc-pVTZ | 0.0048 | 0.3467 | 0.0733 | 0.0009 | 0.0733 | 0.0009 | 0.3467 | 0.0048 | 0.0733 | 0.0009 | 0.0733 | 0.0009 | 0.3515 | 0.0742 | 0.0742 | 0.3515 | 0.0742 | 0.0742 |
| DFT(B3LYP) | cc-pVQZ | 0.0064 | 0.3430 | 0.0740 | 0.0013 | 0.0740 | 0.0013 | 0.3430 | 0.0064 | 0.0740 | 0.0013 | 0.0740 | 0.0013 | 0.3494 | 0.0753 | 0.0753 | 0.3494 | 0.0753 | 0.0753 |
| DFT(M06-2X) | 6-31G(d,p) | 0.0008 | 0.3591 | 0.0699 | 0.0001 | 0.0699 | 0.0001 | 0.3591 | 0.0008 | 0.0699 | 0.0001 | 0.0699 | 0.0001 | 0.3599 | 0.0701 | 0.0701 | 0.3599 | 0.0701 | 0.0701 |
| DFT(M06-2X) | cc-pVDZ | 0.0018 | 0.3544 | 0.0715 | 0.0003 | 0.0715 | 0.0003 | 0.3544 | 0.0018 | 0.0715 | 0.0003 | 0.0715 | 0.0003 | 0.3562 | 0.0719 | 0.0719 | 0.3562 | 0.0719 | 0.0719 |
| DFT(M06-2X) | cc-pVTZ | 0.0039 | 0.3448 | 0.0749 | 0.0008 | 0.0749 | 0.0008 | 0.3448 | 0.0039 | 0.0749 | 0.0008 | 0.0749 | 0.0008 | 0.3487 | 0.0756 | 0.0756 | 0.3487 | 0.0756 | 0.0756 |
| DFT(M06-2X) | cc-pVQZ | 0.0046 | 0.3403 | 0.0766 | 0.0010 | 0.0766 | 0.0010 | 0.3403 | 0.0046 | 0.0766 | 0.0010 | 0.0766 | 0.0010 | 0.3449 | 0.0776 | 0.0776 | 0.3449 | 0.0776 | 0.0776 |
| DFT(CAM-B3LYP) | 6-31G(d,p) | 0.0009 | 0.3608 | 0.0690 | 0.0002 | 0.0690 | 0.0002 | 0.3608 | 0.0009 | 0.0690 | 0.0002 | 0.0690 | 0.0002 | 0.3617 | 0.0691 | 0.0691 | 0.3617 | 0.0691 | 0.0691 |
| DFT(CAM-B3LYP) | cc-pVDZ | 0.0020 | 0.3562 | 0.0705 | 0.0004 | 0.0705 | 0.0004 | 0.3562 | 0.0020 | 0.0705 | 0.0004 | 0.0705 | 0.0004 | 0.3583 | 0.0709 | 0.0709 | 0.3583 | 0.0709 | 0.0709 |
| DFT(CAM-B3LYP) | cc-pVTZ | 0.0045 | 0.3464 | 0.0737 | 0.0009 | 0.0737 | 0.0009 | 0.3464 | 0.0045 | 0.0737 | 0.0009 | 0.0737 | 0.0009 | 0.3509 | 0.0745 | 0.0745 | 0.3509 | 0.0745 | 0.0745 |
| DFT(CAM-B3LYP) | cc-pVQZ | 0.0058 | 0.3430 | 0.0745 | 0.0011 | 0.0745 | 0.0011 | 0.3430 | 0.0058 | 0.0745 | 0.0011 | 0.0745 | 0.0011 | 0.3488 | 0.0756 | 0.0756 | 0.3488 | 0.0756 | 0.0756 |

**Table S22.** Mulliken atomic spin densities obtained with chosen computational approaches for antiquinoid isomer of cationic form of benzene molecule. In the right part of the table atomic spin densities of hydrogens are summed into heavy atoms they are connected to. Please note, that data for CASPT2 method are computed for equilibrium geometries obtained at this level of theory, however spin densities are computed basing on CASSCF wavefunction.

BZ(AQ) R+

| Method | Basis set | H1 | C2 | C3 | H4 | C5 | H6 | C7 | H8 | C9 | H10 | C11 | H12 | H1+C2 | C3+H4 | C5+H6 | C7+H8 | C9+J10 | C11+H12 |
|---|---|---|---|---|---|---|---|---|---|---|---|---|---|---|---|---|---|---|---|
| CASPT2, 0-IPEA | 6-31G(d,p) | 0.0006 | 0.2986 | -0.0990 | -0.0002 | 0.2993 | 0.0006 | 0.2986 | 0.0006 | -0.0990 | -0.0002 | 0.2993 | 0.0006 | 0.2992 | -0.0992 | 0.2999 | 0.2992 | -0.0992 | 0.2999 |
| CASPT2, 0-IPEA | cc-pVDZ | 0.0015 | 0.2963 | -0.0953 | -0.0004 | 0.2964 | 0.0015 | 0.2963 | 0.0015 | -0.0953 | -0.0004 | 0.2964 | 0.0015 | 0.2978 | -0.0957 | 0.2979 | 0.2978 | -0.0957 | 0.2979 |
| CASPT2, 0-IPEA | cc-pVTZ | 0.0031 | 0.2896 | -0.0846 | -0.0008 | 0.2896 | 0.0031 | 0.2896 | 0.0031 | -0.0846 | -0.0008 | 0.2896 | 0.0031 | 0.2927 | -0.0854 | 0.2927 | 0.2927 | -0.0854 | 0.2927 |
| CASPT2, S-IPEA | 6-31G(d,p) | 0.0006 | 0.2988 | -0.0989 | -0.0002 | 0.2990 | 0.0006 | 0.2988 | 0.0006 | -0.0989 | -0.0002 | 0.2990 | 0.0006 | 0.2994 | -0.0991 | 0.2996 | 0.2994 | -0.0991 | 0.2996 |
| CASPT2, S-IPEA | cc-pVDZ | 0.0015 | 0.2959 | -0.0951 | -0.0004 | 0.2967 | 0.0015 | 0.2959 | 0.0015 | -0.0951 | -0.0004 | 0.2967 | 0.0015 | 0.2974 | -0.0955 | 0.2982 | 0.2974 | -0.0955 | 0.2982 |
| CASPT2, S-IPEA | cc-pVTZ | 0.0031 | 0.2894 | -0.0844 | -0.0008 | 0.2896 | 0.0031 | 0.2894 | 0.0031 | -0.0844 | -0.0008 | 0.2896 | 0.0031 | 0.2925 | -0.0852 | 0.2927 | 0.2925 | -0.0852 | 0.2927 |
| CASSCF | 6-31G(d,p) | 0.0007 | 0.2987 | -0.0985 | -0.0002 | 0.2987 | 0.0007 | 0.2987 | 0.0007 | -0.0985 | -0.0002 | 0.2987 | 0.0007 | 0.2994 | -0.0987 | 0.2994 | 0.2994 | -0.0987 | 0.2994 |
| CASSCF | cc-pVDZ | 0.0015 | 0.2956 | -0.0939 | -0.0004 | 0.2956 | 0.0015 | 0.2956 | 0.0015 | -0.0939 | -0.0004 | 0.2956 | 0.0015 | 0.2971 | -0.0943 | 0.2971 | 0.2971 | -0.0943 | 0.2971 |
| CASSCF | cc-pVTZ | 0.0031 | 0.2893 | -0.0840 | -0.0008 | 0.2893 | 0.0031 | 0.2893 | 0.0031 | -0.0840 | -0.0008 | 0.2893 | 0.0031 | 0.2924 | -0.0848 | 0.2924 | 0.2924 | -0.0848 | 0.2924 |
| CASSCF | cc-pVQZ | 0.0038 | 0.2868 | -0.0803 | -0.0010 | 0.2868 | 0.0038 | 0.2868 | 0.0038 | -0.0803 | -0.0010 | 0.2868 | 0.0038 | 0.2906 | -0.0813 | 0.2906 | 0.2906 | -0.0813 | 0.2906 |
| DFT(B3LYP) | 6-31G(d,p) | 0.0007 | 0.2478 | 0.0030 | 0.0000 | 0.2478 | 0.0007 | 0.2478 | 0.0007 | 0.0030 | 0.0000 | 0.2478 | 0.0007 | 0.2485 | 0.0030 | 0.2485 | 0.2485 | 0.0030 | 0.2485 |
| DFT(B3LYP) | cc-pVDZ | 0.0015 | 0.2453 | 0.0063 | 0.0000 | 0.2453 | 0.0015 | 0.2453 | 0.0015 | 0.0063 | 0.0000 | 0.2453 | 0.0015 | 0.2468 | 0.0063 | 0.2468 | 0.2468 | 0.0063 | 0.2468 |
| DFT(B3LYP) | cc-pVTZ | 0.0034 | 0.2405 | 0.0122 | 0.0000 | 0.2405 | 0.0034 | 0.2405 | 0.0034 | 0.0122 | 0.0000 | 0.2405 | 0.0034 | 0.2439 | 0.0122 | 0.2439 | 0.2439 | 0.0122 | 0.2439 |
| DFT(B3LYP) | cc-pVQZ | 0.0045 | 0.2387 | 0.0136 | 0.0000 | 0.2387 | 0.0045 | 0.2387 | 0.0045 | 0.0136 | 0.0000 | 0.2387 | 0.0045 | 0.2432 | 0.0136 | 0.2432 | 0.2432 | 0.0136 | 0.2432 |
| DFT(M06-2X) | 6-31G(d,p) | 0.0005 | 0.2480 | 0.0030 | 0.0000 | 0.2480 | 0.0005 | 0.2480 | 0.0005 | 0.0030 | 0.0000 | 0.2480 | 0.0005 | 0.2485 | 0.0030 | 0.2485 | 0.2485 | 0.0030 | 0.2485 |
| DFT(M06-2X) | cc-pVDZ | 0.0013 | 0.2455 | 0.0065 | 0.0000 | 0.2455 | 0.0013 | 0.2455 | 0.0013 | 0.0065 | 0.0000 | 0.2455 | 0.0013 | 0.2468 | 0.0065 | 0.2468 | 0.2468 | 0.0065 | 0.2468 |
| DFT(M06-2X) | cc-pVTZ | 0.0028 | 0.2409 | 0.0127 | 0.0000 | 0.2409 | 0.0028 | 0.2409 | 0.0028 | 0.0127 | 0.0000 | 0.2409 | 0.0028 | 0.2436 | 0.0127 | 0.2436 | 0.2436 | 0.0127 | 0.2436 |
| DFT(M06-2X) | cc-pVQZ | 0.0033 | 0.2389 | 0.0155 | 0.0000 | 0.2389 | 0.0033 | 0.2389 | 0.0033 | 0.0155 | 0.0000 | 0.2389 | 0.0033 | 0.2423 | 0.0155 | 0.2423 | 0.2423 | 0.0155 | 0.2423 |
| DFT(CAM-B3LYP) | 6-31G(d,p) | 0.0006 | 0.2478 | 0.0031 | 0.0000 | 0.2478 | 0.0006 | 0.2478 | 0.0006 | 0.0031 | 0.0000 | 0.2478 | 0.0006 | 0.2485 | 0.0031 | 0.2485 | 0.2485 | 0.0031 | 0.2485 |
| DFT(CAM-B3LYP) | cc-pVDZ | 0.0014 | 0.2454 | 0.0065 | 0.0000 | 0.2454 | 0.0014 | 0.2454 | 0.0014 | 0.0065 | 0.0000 | 0.2454 | 0.0014 | 0.2468 | 0.0065 | 0.2468 | 0.2468 | 0.0065 | 0.2468 |
| DFT(CAM-B3LYP) | cc-pVTZ | 0.0031 | 0.2406 | 0.0125 | 0.0000 | 0.2406 | 0.0031 | 0.2406 | 0.0031 | 0.0125 | 0.0000 | 0.2406 | 0.0031 | 0.2437 | 0.0125 | 0.2437 | 0.2437 | 0.0125 | 0.2437 |
| DFT(CAM-B3LYP) | cc-pVQZ | 0.0040 | 0.2389 | 0.0140 | 0.0000 | 0.2389 | 0.0040 | 0.2389 | 0.0040 | 0.0140 | 0.0000 | 0.2389 | 0.0040 | 0.2430 | 0.0141 | 0.2430 | 0.2430 | 0.0141 | 0.2430 |

**Table S23.** Mulliken atomic spin densities obtained with chosen computational approaches for quinoid isomer of benzene molecule in the first electronic triplet excided state. In the right part of the table atomic spin densities of hydrogens are summed into heavy atoms they are connected to. Please note, that data for CASPT2 method are computed for equilibrium geometries obtained at this level of theory, however spin densities are computed basing on CASSCF wavefunction.

$$BZ(Q)\ R^0_{T1}$$

| Method | Basis set | H1 | C2 | C3 | H4 | C5 | H6 | C7 | H8 | C9 | H10 | C11 | H12 | H1+C2 | C3+H4 | C5+H6 | C7+H8 | C9+J10 | C11+H12 |
|---|---|---|---|---|---|---|---|---|---|---|---|---|---|---|---|---|---|---|---|
| CASPT2, 0-IPEA | 6-31G(d,p) | 0.0027 | 0.7457 | 0.1253 | 0.0005 | 0.1253 | 0.0005 | 0.7457 | 0.0027 | 0.1253 | 0.0005 | 0.1253 | 0.0005 | 0.7484 | 0.1258 | 0.1258 | 0.7484 | 0.1258 | 0.1258 |
| CASPT2, 0-IPEA | cc-pVDZ | 0.0054 | 0.7439 | 0.1244 | 0.0010 | 0.1244 | 0.0010 | 0.7439 | 0.0054 | 0.1244 | 0.0010 | 0.1244 | 0.0010 | 0.7493 | 0.1254 | 0.1254 | 0.7493 | 0.1254 | 0.1254 |
| CASPT2, 0-IPEA | cc-pVTZ | 0.0112 | 0.7494 | 0.1178 | 0.0019 | 0.1178 | 0.0019 | 0.7494 | 0.0112 | 0.1178 | 0.0019 | 0.1178 | 0.0019 | 0.7606 | 0.1197 | 0.1197 | 0.7606 | 0.1197 | 0.1197 |
| CASPT2, S-IPEA | 6-31G(d,p) | 0.0028 | 0.7654 | 0.1155 | 0.0004 | 0.1155 | 0.0004 | 0.7654 | 0.0028 | 0.1155 | 0.0004 | 0.1155 | 0.0004 | 0.7682 | 0.1159 | 0.1159 | 0.7682 | 0.1159 | 0.1159 |
| CASPT2, S-IPEA | cc-pVDZ | 0.0055 | 0.7639 | 0.1144 | 0.0009 | 0.1144 | 0.0009 | 0.7639 | 0.0055 | 0.1144 | 0.0009 | 0.1144 | 0.0009 | 0.7694 | 0.1153 | 0.1153 | 0.7694 | 0.1153 | 0.1153 |
| CASPT2, S-IPEA | cc-pVTZ | 0.0114 | 0.7638 | 0.1106 | 0.0018 | 0.1106 | 0.0018 | 0.7638 | 0.0114 | 0.1106 | 0.0018 | 0.1106 | 0.0018 | 0.7752 | 0.1124 | 0.1124 | 0.7752 | 0.1124 | 0.1124 |
| CASSCF | 6-31G(d,p) | 0.0029 | 0.7751 | 0.1106 | 0.0004 | 0.1106 | 0.0004 | 0.7751 | 0.0029 | 0.1106 | 0.0004 | 0.1106 | 0.0004 | 0.7780 | 0.1110 | 0.1110 | 0.7780 | 0.1110 | 0.1110 |
| CASSCF | cc-pVDZ | 0.0058 | 0.7732 | 0.1096 | 0.0009 | 0.1096 | 0.0009 | 0.7732 | 0.0058 | 0.1096 | 0.0009 | 0.1096 | 0.0009 | 0.7790 | 0.1105 | 0.1105 | 0.7790 | 0.1105 | 0.1105 |
| CASSCF | cc-pVTZ | 0.0117 | 0.7718 | 0.1065 | 0.0018 | 0.1065 | 0.0018 | 0.7718 | 0.0117 | 0.1065 | 0.0018 | 0.1065 | 0.0018 | 0.7835 | 0.1083 | 0.1083 | 0.7835 | 0.1083 | 0.1083 |
| CASSCF | cc-pVQZ | 0.0150 | 0.7697 | 0.1053 | 0.0023 | 0.1053 | 0.0023 | 0.7697 | 0.0150 | 0.1053 | 0.0023 | 0.1053 | 0.0023 | 0.7847 | 0.1076 | 0.1076 | 0.7847 | 0.1076 | 0.1076 |
| DFT(B3LYP) | 6-31G(d,p) | -0.0430 | 0.8976 | 0.0805 | -0.0078 | 0.0805 | -0.0078 | 0.8976 | -0.0430 | 0.0805 | -0.0078 | 0.0805 | -0.0078 | 0.8546 | 0.0727 | 0.0727 | 0.8546 | 0.0727 | 0.0727 |
| DFT(B3LYP) | cc-pVDZ | -0.0382 | 0.8662 | 0.0931 | -0.0071 | 0.0931 | -0.0071 | 0.8662 | -0.0382 | 0.0931 | -0.0071 | 0.0931 | -0.0071 | 0.8280 | 0.0860 | 0.0860 | 0.8280 | 0.0860 | 0.0860 |
| DFT(B3LYP) | cc-pVTZ | -0.0318 | 0.8397 | 0.1029 | -0.0068 | 0.1029 | -0.0068 | 0.8397 | -0.0318 | 0.1029 | -0.0068 | 0.1029 | -0.0068 | 0.8079 | 0.0960 | 0.0960 | 0.8079 | 0.0960 | 0.0960 |
| DFT(B3LYP) | cc-pVQZ | -0.0360 | 0.8508 | 0.1017 | -0.0091 | 0.1017 | -0.0091 | 0.8508 | -0.0360 | 0.1017 | -0.0091 | 0.1017 | -0.0091 | 0.8148 | 0.0926 | 0.0926 | 0.8148 | 0.0926 | 0.0926 |
| DFT(M06-2X) | 6-31G(d,p) | -0.0482 | 0.9436 | 0.0610 | -0.0087 | 0.0610 | -0.0087 | 0.9436 | -0.0482 | 0.0610 | -0.0087 | 0.0610 | -0.0087 | 0.8954 | 0.0523 | 0.0523 | 0.8954 | 0.0523 | 0.0523 |
| DFT(M06-2X) | cc-pVDZ | -0.0399 | 0.9061 | 0.0743 | -0.0074 | 0.0743 | -0.0074 | 0.9061 | -0.0399 | 0.0743 | -0.0074 | 0.0743 | -0.0074 | 0.8662 | 0.0669 | 0.0669 | 0.8662 | 0.0669 | 0.0669 |
| DFT(M06-2X) | cc-pVTZ | -0.0443 | 0.9097 | 0.0754 | -0.0081 | 0.0754 | -0.0081 | 0.9097 | -0.0443 | 0.0754 | -0.0081 | 0.0754 | -0.0081 | 0.8655 | 0.0673 | 0.0673 | 0.8655 | 0.0673 | 0.0673 |
| DFT(M06-2X) | cc-pVQZ | 0.0074 | 0.8071 | 0.0800 | 0.0127 | 0.0800 | 0.0127 | 0.8071 | 0.0074 | 0.0800 | 0.0127 | 0.0800 | 0.0127 | 0.8145 | 0.0927 | 0.0927 | 0.8145 | 0.0927 | 0.0927 |
| DFT(CAM-B3LYP) | 6-31G(d,p) | -0.0441 | 0.9274 | 0.0657 | -0.0073 | 0.0657 | -0.0073 | 0.9274 | -0.0441 | 0.0657 | -0.0073 | 0.0657 | -0.0073 | 0.8833 | 0.0584 | 0.0584 | 0.8833 | 0.0584 | 0.0584 |
| DFT(CAM-B3LYP) | cc-pVDZ | -0.0388 | 0.8940 | 0.0789 | -0.0066 | 0.0789 | -0.0066 | 0.8940 | -0.0388 | 0.0789 | -0.0066 | 0.0789 | -0.0066 | 0.8552 | 0.0724 | 0.0724 | 0.8552 | 0.0724 | 0.0724 |
| DFT(CAM-B3LYP) | cc-pVTZ | -0.0346 | 0.8692 | 0.0890 | -0.0063 | 0.0890 | -0.0063 | 0.8692 | -0.0346 | 0.0890 | -0.0063 | 0.0890 | -0.0063 | 0.8346 | 0.0827 | 0.0827 | 0.8346 | 0.0827 | 0.0827 |
| DFT(CAM-B3LYP) | cc-pVQZ | -0.0423 | 0.8814 | 0.0884 | -0.0080 | 0.0884 | -0.0080 | 0.8814 | -0.0423 | 0.0884 | -0.0080 | 0.0884 | -0.0080 | 0.8391 | 0.0804 | 0.0804 | 0.8391 | 0.0804 | 0.0804 |

**Table S24.** Mulliken atomic spin densities obtained with chosen computational approaches for antiquinoid isomer of benzene molecule in the first electronic triplet excided state. In the right part of the table atomic spin densities of hydrogens are summed into heavy atoms they are connected to. Please note, that data for CASPT2 method are computed for equilibrium geometries obtained at this level of theory, however spin densities are computed basing on CASSCF wavefunction.

$$BZ(AQ)\ R^0_{T1}$$

| Method | Basis set | H1 | C2 | C3 | H4 | C5 | H6 | C7 | H8 | C9 | H10 | C11 | H12 | H1+C2 | C3+H4 | C5+H6 | C7+H8 | C9+J10 | C11+H12 |
|---|---|---|---|---|---|---|---|---|---|---|---|---|---|---|---|---|---|---|---|
| CASPT2, 0-IPEA | 6-31G(d,p) | 0.0019 | 0.5260 | -0.0556 | -0.0001 | 0.5260 | 0.0019 | 0.5260 | 0.0019 | -0.0556 | -0.0001 | 0.5260 | 0.0019 | 0.5279 | -0.0557 | 0.5279 | 0.5279 | -0.0557 | 0.5279 |
| CASPT2, 0-IPEA | cc-pVDZ | 0.0037 | 0.5249 | -0.0570 | -0.0002 | 0.5249 | 0.0037 | 0.5249 | 0.0037 | -0.0570 | -0.0002 | 0.5249 | 0.0037 | 0.5286 | -0.0572 | 0.5286 | 0.5286 | -0.0572 | 0.5286 |
| CASPT2, 0-IPEA | cc-pVTZ | 0.0077 | 0.5275 | -0.0699 | -0.0007 | 0.5275 | 0.0077 | 0.5275 | 0.0077 | -0.0699 | -0.0007 | 0.5275 | 0.0077 | 0.5352 | -0.0706 | 0.5352 | 0.5352 | -0.0706 | 0.5352 |
| CASPT2, S-IPEA | 6-31G(d,p) | 0.0019 | 0.5377 | -0.0789 | -0.0002 | 0.5377 | 0.0019 | 0.5377 | 0.0019 | -0.0789 | -0.0002 | 0.5377 | 0.0019 | 0.5396 | -0.0791 | 0.5396 | 0.5396 | -0.0791 | 0.5396 |
| CASPT2, S-IPEA | cc-pVDZ | 0.0038 | 0.5363 | -0.0798 | -0.0004 | 0.5363 | 0.0038 | 0.5363 | 0.0038 | -0.0798 | -0.0004 | 0.5363 | 0.0038 | 0.5401 | -0.0802 | 0.5401 | 0.5401 | -0.0802 | 0.5401 |
| CASPT2, S-IPEA | cc-pVTZ | 0.0078 | 0.5341 | -0.0830 | -0.0008 | 0.5341 | 0.0078 | 0.5341 | 0.0078 | -0.0830 | -0.0008 | 0.5341 | 0.0078 | 0.5419 | -0.0838 | 0.5419 | 0.5419 | -0.0838 | 0.5419 |
| CASSCF | 6-31G(d,p) | 0.0019 | 0.5383 | -0.0803 | -0.0002 | 0.5383 | 0.0019 | 0.5383 | 0.0019 | -0.0803 | -0.0002 | 0.5383 | 0.0019 | 0.5402 | -0.0805 | 0.5402 | 0.5402 | -0.0805 | 0.5402 |
| CASSCF | cc-pVDZ | 0.0040 | 0.5370 | -0.0815 | -0.0004 | 0.5370 | 0.0040 | 0.5370 | 0.0040 | -0.0815 | -0.0004 | 0.5370 | 0.0040 | 0.5410 | -0.0819 | 0.5410 | 0.5410 | -0.0819 | 0.5410 |
| CASSCF | cc-pVTZ | 0.0079 | 0.5341 | -0.0832 | -0.0009 | 0.5341 | 0.0079 | 0.5341 | 0.0079 | -0.0832 | -0.0009 | 0.5341 | 0.0079 | 0.5420 | -0.0841 | 0.5420 | 0.5420 | -0.0841 | 0.5420 |
| CASSCF | cc-pVQZ | 0.0102 | 0.5316 | -0.0826 | -0.0009 | 0.5316 | 0.0102 | 0.5316 | 0.0102 | -0.0826 | -0.0009 | 0.5316 | 0.0102 | 0.5418 | -0.0835 | 0.5418 | 0.5418 | -0.0835 | 0.5418 |
| DFT(B3LYP) | 6-31G(d,p) | -0.0330 | 0.6613 | -0.2640 | 0.0073 | 0.6613 | -0.0330 | 0.6613 | -0.0330 | -0.2640 | 0.0073 | 0.6613 | -0.0330 | 0.6283 | -0.2567 | 0.6283 | 0.6283 | -0.2567 | 0.6283 |
| DFT(B3LYP) | cc-pVDZ | -0.0295 | 0.6392 | -0.2258 | 0.0064 | 0.6392 | -0.0295 | 0.6392 | -0.0295 | -0.2258 | 0.0064 | 0.6392 | -0.0295 | 0.6097 | -0.2194 | 0.6097 | 0.6097 | -0.2194 | 0.6097 |
| DFT(B3LYP) | cc-pVTZ | -0.0252 | 0.6175 | -0.1892 | 0.0046 | 0.6175 | -0.0252 | 0.6175 | -0.0252 | -0.1892 | 0.0046 | 0.6175 | -0.0252 | 0.5923 | -0.1847 | 0.5923 | 0.5923 | -0.1847 | 0.5923 |
| DFT(B3LYP) | cc-pVQZ | -0.0288 | 0.6258 | -0.1972 | 0.0032 | 0.6258 | -0.0288 | 0.6258 | -0.0288 | -0.1972 | 0.0032 | 0.6258 | -0.0288 | 0.5970 | -0.1940 | 0.5970 | 0.5970 | -0.1940 | 0.5970 |
| DFT(M06-2X) | 6-31G(d,p) | -0.0353 | 0.6680 | -0.2712 | 0.0059 | 0.6680 | -0.0353 | 0.6680 | -0.0353 | -0.2712 | 0.0059 | 0.6680 | -0.0353 | 0.6327 | -0.2654 | 0.6327 | 0.6327 | -0.2654 | 0.6327 |
| DFT(M06-2X) | cc-pVDZ | -0.0292 | 0.6531 | -0.2522 | 0.0045 | 0.6531 | -0.0292 | 0.6531 | -0.0292 | -0.2522 | 0.0045 | 0.6531 | -0.0292 | 0.6239 | -0.2477 | 0.6239 | 0.6239 | -0.2477 | 0.6239 |
| DFT(M06-2X) | cc-pVTZ | -0.0385 | 0.6796 | -0.2982 | 0.0160 | 0.6796 | -0.0385 | 0.6796 | -0.0385 | -0.2982 | 0.0160 | 0.6796 | -0.0385 | 0.6411 | -0.2822 | 0.6411 | 0.6411 | -0.2822 | 0.6411 |
| DFT(M06-2X) | cc-pVQZ | 0.0036 | 0.6017 | -0.2264 | 0.0158 | 0.6017 | 0.0036 | 0.6017 | 0.0036 | -0.2264 | 0.0158 | 0.6017 | 0.0036 | 0.6053 | -0.2107 | 0.6053 | 0.6053 | -0.2107 | 0.6053 |
| DFT(CAM-B3LYP) | 6-31G(d,p) | -0.0340 | 0.6870 | -0.3152 | 0.0091 | 0.6870 | -0.0340 | 0.6870 | -0.0340 | -0.3152 | 0.0091 | 0.6870 | -0.0340 | 0.6530 | -0.3061 | 0.6530 | 0.6530 | -0.3061 | 0.6530 |
| DFT(CAM-B3LYP) | cc-pVDZ | -0.0300 | 0.6627 | -0.2733 | 0.0080 | 0.6627 | -0.0300 | 0.6627 | -0.0300 | -0.2733 | 0.0080 | 0.6627 | -0.0300 | 0.6327 | -0.2653 | 0.6327 | 0.6327 | -0.2653 | 0.6327 |
| DFT(CAM-B3LYP) | cc-pVTZ | -0.0271 | 0.6408 | -0.2342 | 0.0067 | 0.6408 | -0.0271 | 0.6408 | -0.0271 | -0.2342 | 0.0067 | 0.6408 | -0.0271 | 0.6137 | -0.2274 | 0.6137 | 0.6137 | -0.2274 | 0.6137 |
| DFT(CAM-B3LYP) | cc-pVQZ | -0.0328 | 0.6487 | -0.2387 | 0.0070 | 0.6487 | -0.0328 | 0.6487 | -0.0328 | -0.2387 | 0.0070 | 0.6487 | -0.0328 | 0.6159 | -0.2317 | 0.6159 | 0.6159 | -0.2317 | 0.6159 |

**Table S25.** Mulliken atomic spin densities for benzene molecule in the ground electronic state predicted as combination ($R^- + R^+ - R^0_{T1}$) of atomic spin densities computed for other electronic states structures in their quinoid variants. In the right part of the table there are data for summed spin densities of heavy atoms and hydrogens connected to them. Please note, that data for CASPT2 method are computed for equilibrium geometries obtained at this level of theory, however spin densities are computed basing on CASSCF wavefunction.

$$BZ(Q) \quad R^- + R^+ - R^0_{T1}$$

| Method | Basis set | H1 | C2 | C3 | H4 | C5 | H6 | C7 | H8 | C9 | H10 | C11 | H12 | H1+C2 | C3+H4 | C5+H6 | C7+H8 | C9+J10 | C11+H12 |
|---|---|---|---|---|---|---|---|---|---|---|---|---|---|---|---|---|---|---|---|
| CASPT2, 0-IPEA | 6-31G(d,p) | 0.0003 | 0.1153 | -0.0583 | -0.0001 | -0.0571 | -0.0001 | 0.1153 | 0.0003 | -0.0583 | -0.0001 | -0.0571 | -0.0001 | 0.1156 | -0.0584 | -0.0572 | 0.1156 | -0.0584 | -0.0572 |
| CASPT2, 0-IPEA | cc-pVDZ | 0.0007 | 0.1055 | -0.0529 | -0.0003 | -0.0529 | -0.0003 | 0.1055 | 0.0007 | -0.0529 | -0.0003 | -0.0529 | -0.0003 | 0.1062 | -0.0532 | -0.0532 | 0.1062 | -0.0532 | -0.0532 |
| CASPT2, 0-IPEA | cc-pVTZ | 0.0021 | 0.0735 | -0.0376 | -0.0003 | -0.0376 | -0.0003 | 0.0735 | 0.0021 | -0.0376 | -0.0003 | -0.0376 | -0.0003 | 0.0756 | -0.0379 | -0.0379 | 0.0756 | -0.0379 | -0.0379 |
| CASPT2, S-IPEA | 6-31G(d,p) | 0.0002 | 0.0956 | -0.0479 | 0.0000 | -0.0477 | 0.0000 | 0.0956 | 0.0002 | -0.0479 | 0.0000 | -0.0477 | 0.0000 | 0.0958 | -0.0479 | -0.0477 | 0.0958 | -0.0479 | -0.0477 |
| CASPT2, S-IPEA | cc-pVDZ | 0.0006 | 0.0854 | -0.0428 | -0.0002 | -0.0428 | -0.0002 | 0.0854 | 0.0006 | -0.0428 | -0.0002 | -0.0428 | -0.0002 | 0.0860 | -0.0430 | -0.0430 | 0.0860 | -0.0430 | -0.0430 |
| CASPT2, S-IPEA | cc-pVTZ | 0.0019 | 0.0590 | -0.0303 | -0.0002 | -0.0303 | -0.0002 | 0.0590 | 0.0019 | -0.0303 | -0.0002 | -0.0303 | -0.0002 | 0.0609 | -0.0305 | -0.0305 | 0.0609 | -0.0305 | -0.0305 |
| CASSCF | 6-31G(d,p) | 0.0002 | 0.0858 | -0.0430 | 0.0000 | -0.0430 | 0.0000 | 0.0858 | 0.0002 | -0.0430 | 0.0000 | -0.0430 | 0.0000 | 0.0860 | -0.0430 | -0.0430 | 0.0860 | -0.0430 | -0.0430 |
| CASSCF | cc-pVDZ | 0.0005 | 0.0742 | -0.0372 | -0.0002 | -0.0372 | -0.0002 | 0.0742 | 0.0005 | -0.0372 | -0.0002 | -0.0372 | -0.0002 | 0.0747 | -0.0374 | -0.0374 | 0.0747 | -0.0374 | -0.0374 |
| CASSCF | cc-pVTZ | 0.0018 | 0.0506 | -0.0261 | -0.0002 | -0.0261 | -0.0002 | 0.0506 | 0.0018 | -0.0261 | -0.0002 | -0.0261 | -0.0002 | 0.0524 | -0.0263 | -0.0263 | 0.0524 | -0.0263 | -0.0263 |
| CASSCF | cc-pVQZ | 0.0062 | 0.0438 | -0.0256 | 0.0007 | -0.0256 | 0.0007 | 0.0438 | 0.0062 | -0.0256 | 0.0007 | -0.0256 | 0.0007 | 0.0500 | -0.0249 | -0.0249 | 0.0500 | -0.0249 | -0.0249 |
| DFT(B3LYP) | 6-31G(d,p) | 0.0461 | -0.1780 | 0.0575 | 0.0084 | 0.0575 | 0.0084 | -0.1780 | 0.0461 | 0.0575 | 0.0084 | 0.0575 | 0.0084 | -0.1319 | 0.0660 | 0.0660 | -0.1319 | 0.0660 | 0.0660 |
| DFT(B3LYP) | cc-pVDZ | 0.0445 | -0.1580 | 0.0483 | 0.0084 | 0.0483 | 0.0084 | -0.1580 | 0.0445 | 0.0483 | 0.0084 | 0.0483 | 0.0084 | -0.1135 | 0.0567 | 0.0567 | -0.1135 | 0.0567 | 0.0567 |
| DFT(B3LYP) | cc-pVTZ | 0.0463 | -0.1517 | 0.0428 | 0.0099 | 0.0428 | 0.0099 | -0.1517 | 0.0463 | 0.0428 | 0.0099 | 0.0428 | 0.0099 | -0.1053 | 0.0527 | 0.0527 | -0.1053 | 0.0527 | 0.0527 |
| DFT(B3LYP) | cc-pVQZ | 0.0612 | -0.1710 | 0.0404 | 0.0145 | 0.0404 | 0.0145 | -0.1710 | 0.0612 | 0.0404 | 0.0145 | 0.0404 | 0.0145 | -0.1098 | 0.0549 | 0.0549 | -0.1098 | 0.0549 | 0.0549 |
| DFT(M06-2X) | 6-31G(d,p) | 0.0509 | -0.2292 | 0.0799 | 0.0093 | 0.0799 | 0.0093 | -0.2292 | 0.0509 | 0.0799 | 0.0093 | 0.0799 | 0.0093 | -0.1783 | 0.0892 | 0.0892 | -0.1783 | 0.0892 | 0.0892 |
| DFT(M06-2X) | cc-pVDZ | 0.0454 | -0.2033 | 0.0704 | 0.0085 | 0.0704 | 0.0085 | -0.2033 | 0.0454 | 0.0704 | 0.0085 | 0.0704 | 0.0085 | -0.1578 | 0.0789 | 0.0789 | -0.1578 | 0.0789 | 0.0789 |
| DFT(M06-2X) | cc-pVTZ | 0.0566 | -0.2261 | 0.0740 | 0.0107 | 0.0740 | 0.0107 | -0.2261 | 0.0566 | 0.0740 | 0.0107 | 0.0740 | 0.0107 | -0.1694 | 0.0847 | 0.0847 | -0.1694 | 0.0847 | 0.0847 |
| DFT(M06-2X) | cc-pVQZ | 0.0113 | -0.1317 | 0.0687 | -0.0085 | 0.0687 | -0.0085 | -0.1317 | 0.0113 | 0.0687 | -0.0085 | 0.0687 | -0.0085 | -0.1204 | 0.0602 | 0.0602 | -0.1204 | 0.0602 | 0.0602 |
| DFT(CAM-B3LYP) | 6-31G(d,p) | 0.0471 | -0.2091 | 0.0731 | 0.0079 | 0.0731 | 0.0079 | -0.2091 | 0.0471 | 0.0731 | 0.0079 | 0.0731 | 0.0079 | -0.1620 | 0.0810 | 0.0810 | -0.1620 | 0.0810 | 0.0810 |
| DFT(CAM-B3LYP) | cc-pVDZ | 0.0148 | -0.0515 | 0.0148 | 0.0036 | 0.0148 | 0.0036 | -0.0515 | 0.0148 | 0.0148 | 0.0036 | 0.0148 | 0.0036 | -0.0368 | 0.0184 | 0.0184 | -0.0368 | 0.0184 | 0.0184 |
| DFT(CAM-B3LYP) | cc-pVTZ | 0.0482 | -0.1826 | 0.0581 | 0.0091 | 0.0581 | 0.0091 | -0.1826 | 0.0482 | 0.0581 | 0.0091 | 0.0581 | 0.0091 | -0.1344 | 0.0672 | 0.0672 | -0.1344 | 0.0672 | 0.0672 |
| DFT(CAM-B3LYP) | cc-pVQZ | 0.0655 | -0.2024 | 0.0555 | 0.0129 | 0.0555 | 0.0129 | -0.2024 | 0.0655 | 0.0555 | 0.0129 | 0.0555 | 0.0129 | -0.1369 | 0.0684 | 0.0684 | -0.1369 | 0.0684 | 0.0684 |

**Table S26.** Mulliken atomic spin densities for benzene molecule in the ground electronic state predicted as combination ($R^- + R^+ - R^0_{T1}$) of atomic spin densities computed for other electronic states structures in their antiquinoid variants. In the right part of the table there are data for summed spin densities of heavy atoms and hydrogens connected to them. Please note, that data for CASPT2 method are computed for equilibrium geometries obtained at this level of theory, however spin densities are computed basing on CASSCF wavefunction. TBD – to be determined.

$$BZ(AQ) \quad R^- + R^+ - R^0_{T1}$$

| Method | Basis set | H1 | C2 | C3 | H4 | C5 | H6 | C7 | H8 | C9 | H10 | C11 | H12 | H1+C2 | C3+H4 | C5+H6 | C7+H8 | C9+J10 | C11+H12 |
|---|---|---|---|---|---|---|---|---|---|---|---|---|---|---|---|---|---|---|---|
| CASPT2, 0-IPEA | 6-31G(d,p) | 0.0001 | 0.0657 | -0.1322 | -0.0003 | 0.0664 | 0.0001 | 0.0657 | 0.0001 | -0.1322 | -0.0003 | 0.0664 | 0.0001 | 0.0658 | -0.1325 | 0.0665 | 0.0658 | -0.1325 | 0.0665 |
| CASPT2, 0-IPEA | cc-pVDZ | 0.0005 | 0.0608 | -0.1221 | -0.0006 | 0.0609 | 0.0005 | 0.0608 | 0.0005 | -0.1221 | -0.0006 | 0.0609 | 0.0005 | 0.0613 | -0.1227 | 0.0614 | 0.0613 | -0.1227 | 0.0614 |
| CASPT2, 0-IPEA | cc-pVTZ | | | | | | | | | TBD | | | | | | | | | |
| CASPT2, S-IPEA | 6-31G(d,p) | 0.0001 | 0.0541 | -0.1086 | -0.0002 | 0.0543 | 0.0001 | 0.0541 | 0.0001 | -0.1086 | -0.0002 | 0.0543 | 0.0001 | 0.0542 | -0.1088 | 0.0544 | 0.0542 | -0.1088 | 0.0544 |
| CASPT2, S-IPEA | cc-pVDZ | 0.0004 | 0.0489 | -0.0989 | -0.0004 | 0.0497 | 0.0004 | 0.0489 | 0.0004 | -0.0989 | -0.0004 | 0.0497 | 0.0004 | 0.0493 | -0.0993 | 0.0501 | 0.0493 | -0.0993 | 0.0501 |
| CASPT2, S-IPEA | cc-pVTZ | | | | | | | | | TBD | | | | | | | | | |
| CASSCF | 6-31G(d,p) | 0.0002 | 0.0532 | -0.1065 | -0.0002 | 0.0532 | 0.0002 | 0.0532 | 0.0002 | -0.1065 | -0.0002 | 0.0532 | 0.0002 | 0.0534 | -0.1067 | 0.0534 | 0.0534 | -0.1067 | 0.0534 |
| CASSCF | cc-pVDZ | 0.0003 | 0.0471 | -0.0945 | -0.0004 | 0.0471 | 0.0003 | 0.0471 | 0.0003 | -0.0945 | -0.0004 | 0.0471 | 0.0003 | 0.0474 | -0.0949 | 0.0474 | 0.0474 | -0.0949 | 0.0474 |
| CASSCF | cc-pVTZ | 0.0012 | 0.0356 | -0.0730 | -0.0007 | 0.0356 | 0.0012 | 0.0356 | 0.0012 | -0.0730 | -0.0007 | 0.0356 | 0.0012 | 0.0368 | -0.0737 | 0.0368 | 0.0368 | -0.0737 | 0.0368 |
| CASSCF | cc-pVQZ | 0.0040 | 0.0321 | -0.0715 | -0.0008 | 0.0321 | 0.0040 | 0.0321 | 0.0040 | -0.0715 | -0.0008 | 0.0321 | 0.0040 | 0.0361 | -0.0723 | 0.0361 | 0.0361 | -0.0723 | 0.0361 |
| DFT(B3LYP) | 6-31G(d,p) | 0.0352 | -0.1681 | 0.2730 | -0.0073 | -0.1681 | 0.0352 | -0.1681 | 0.0352 | 0.2730 | -0.0073 | -0.1681 | 0.0352 | -0.1329 | 0.2658 | -0.1329 | -0.1329 | 0.2658 | -0.1329 |
| DFT(B3LYP) | cc-pVDZ | 0.0339 | -0.1520 | 0.2426 | -0.0064 | -0.1520 | 0.0339 | -0.1520 | 0.0339 | 0.2426 | -0.0064 | -0.1520 | 0.0339 | -0.1181 | 0.2362 | -0.1181 | -0.1181 | 0.2362 | -0.1181 |
| DFT(B3LYP) | cc-pVTZ | 0.0355 | -0.1411 | 0.2157 | -0.0045 | -0.1411 | 0.0355 | -0.1411 | 0.0355 | 0.2157 | -0.0045 | -0.1411 | 0.0355 | -0.1056 | 0.2112 | -0.1056 | -0.1056 | 0.2112 | -0.1056 |
| DFT(B3LYP) | cc-pVQZ | 0.0467 | -0.1540 | 0.2175 | -0.0030 | -0.1540 | 0.0467 | -0.1540 | 0.0467 | 0.2175 | -0.0030 | -0.1540 | 0.0467 | -0.1073 | 0.2145 | -0.1073 | -0.1073 | 0.2145 | -0.1073 |
| DFT(M06-2X) | 6-31G(d,p) | 0.0372 | -0.1746 | 0.2808 | -0.0059 | -0.1746 | 0.0372 | -0.1746 | 0.0372 | 0.2808 | -0.0059 | -0.1746 | 0.0372 | -0.1375 | 0.2749 | -0.1375 | -0.1375 | 0.2749 | -0.1375 |
| DFT(M06-2X) | cc-pVDZ | 0.0331 | -0.1660 | 0.2702 | -0.0045 | -0.1660 | 0.0331 | -0.1660 | 0.0331 | 0.2702 | -0.0045 | -0.1660 | 0.0331 | -0.1329 | 0.2657 | -0.1329 | -0.1329 | 0.2657 | -0.1329 |
| DFT(M06-2X) | cc-pVTZ | 0.0473 | -0.2031 | 0.3275 | -0.0160 | -0.2031 | 0.0473 | -0.2031 | 0.0473 | 0.3275 | -0.0160 | -0.2031 | 0.0473 | -0.1558 | 0.3115 | -0.1558 | -0.1558 | 0.3115 | -0.1558 |
| DFT(M06-2X) | cc-pVQZ | 0.0099 | -0.1302 | 0.2562 | -0.0156 | -0.1302 | 0.0099 | -0.1302 | 0.0099 | 0.2562 | -0.0156 | -0.1302 | 0.0099 | -0.1203 | 0.2406 | -0.1203 | -0.1203 | 0.2406 | -0.1203 |
| DFT(CAM-B3LYP) | 6-31G(d,p) | 0.0361 | -0.1939 | 0.3248 | -0.0091 | -0.1939 | 0.0361 | -0.1939 | 0.0361 | 0.3248 | -0.0091 | -0.1939 | 0.0361 | -0.1579 | 0.3157 | -0.1579 | -0.1579 | 0.3157 | -0.1579 |
| DFT(CAM-B3LYP) | cc-pVDZ | 0.0129 | -0.0869 | 0.1517 | -0.0035 | -0.0869 | 0.0129 | -0.0869 | 0.0129 | 0.1517 | -0.0035 | -0.0869 | 0.0129 | -0.0741 | 0.1482 | -0.0741 | -0.0741 | 0.1482 | -0.0741 |
| DFT(CAM-B3LYP) | cc-pVTZ | 0.0368 | -0.1648 | 0.2627 | -0.0067 | -0.1648 | 0.0368 | -0.1648 | 0.0368 | 0.2627 | -0.0067 | -0.1648 | 0.0368 | -0.1280 | 0.2561 | -0.1280 | -0.1280 | 0.2561 | -0.1280 |
| DFT(CAM-B3LYP) | cc-pVQZ | 0.0493 | -0.1773 | 0.2627 | -0.0068 | -0.1773 | 0.0493 | -0.1773 | 0.0493 | 0.2627 | -0.0068 | -0.1773 | 0.0493 | -0.1280 | 0.2559 | -0.1280 | -0.1280 | 0.2559 | -0.1280 |

Table S27. Mulliken atomic spin densities obtained with chosen computational approaches for anionic form of cyclobutadiene. In the right part of the table atomic spin densities of hydrogens are summed into heavy atoms they are connected to. Please note, that data for CASPT2 method are computed for equilibrium geometries obtained at this level of theory, however spin densities are computed basing on CASSCF wavefunction.

## CBDE R⁻

| Method | Basis set | C1 | H2 | C3 | H4 | C5 | H6 | C7 | H8 | C1+H2 | C3+H4 | C5+H6 | C7+H8 |
|---|---|---|---|---|---|---|---|---|---|---|---|---|---|
| CASPT2, 0-IPEA | 6-31G(d,p) | 0.2511 | 0.0011 | 0.2468 | 0.0011 | 0.2511 | 0.0011 | 0.2468 | 0.0011 | 0.2522 | 0.2479 | 0.2522 | 0.2479 |
| CASPT2, 0-IPEA | cc-pVDZ | 0.2493 | 0.0022 | 0.2465 | 0.0021 | 0.2492 | 0.0022 | 0.2465 | 0.0021 | 0.2515 | 0.2486 | 0.2514 | 0.2486 |
| CASPT2, 0-IPEA | cc-pVTZ | 0.2440 | 0.0049 | 0.2462 | 0.0049 | 0.2440 | 0.0049 | 0.2462 | 0.0049 | 0.2489 | 0.2511 | 0.2489 | 0.2511 |
| CASPT2, S-IPEA | 6-31G(d,p) | 0.2490 | 0.0011 | 0.2488 | 0.0011 | 0.2490 | 0.0011 | 0.2488 | 0.0011 | 0.2501 | 0.2499 | 0.2501 | 0.2499 |
| CASPT2, S-IPEA | cc-pVDZ | 0.2484 | 0.0022 | 0.2473 | 0.0022 | 0.2484 | 0.0022 | 0.2473 | 0.0022 | 0.2506 | 0.2495 | 0.2506 | 0.2495 |
| CASPT2, S-IPEA | cc-pVTZ | 0.2455 | 0.0049 | 0.2447 | 0.0049 | 0.2453 | 0.0049 | 0.2448 | 0.0049 | 0.2504 | 0.2496 | 0.2502 | 0.2497 |
| CASSCF | 6-31G(d,p) | 0.2489 | 0.0011 | 0.2489 | 0.0011 | 0.2489 | 0.0011 | 0.2489 | 0.0011 | 0.2500 | 0.2500 | 0.2500 | 0.2500 |
| CASSCF | cc-pVDZ | 0.2478 | 0.0022 | 0.2478 | 0.0022 | 0.2478 | 0.0022 | 0.2478 | 0.0022 | 0.2500 | 0.2500 | 0.2500 | 0.2500 |
| CASSCF | cc-pVTZ | 0.2448 | 0.0050 | 0.2452 | 0.0050 | 0.2448 | 0.0050 | 0.2452 | 0.0050 | 0.2498 | 0.2502 | 0.2498 | 0.2502 |
| CASSCF | cc-pVQZ | 0.2402 | 0.0097 | 0.2405 | 0.0097 | 0.2402 | 0.0097 | 0.2405 | 0.0097 | 0.2499 | 0.2502 | 0.2499 | 0.2502 |
| DFT(B3LYP) | 6-31G(d,p) | 0.2652 | -0.0152 | 0.2652 | -0.0152 | 0.2652 | -0.0152 | 0.2652 | -0.0152 | 0.2500 | 0.2500 | 0.2500 | 0.2500 |
| DFT(B3LYP) | cc-pVDZ | 0.2644 | -0.0144 | 0.2644 | -0.0144 | 0.2644 | -0.0144 | 0.2644 | -0.0144 | 0.2500 | 0.2500 | 0.2500 | 0.2500 |
| DFT(B3LYP) | cc-pVTZ | 0.2613 | -0.0112 | 0.2613 | -0.0112 | 0.2612 | -0.0112 | 0.2612 | -0.0112 | 0.2500 | 0.2500 | 0.2500 | 0.2500 |
| DFT(B3LYP) | cc-pVQZ | 0.2605 | -0.0105 | 0.2605 | -0.0105 | 0.2605 | -0.0105 | 0.2605 | -0.0105 | 0.2500 | 0.2500 | 0.2500 | 0.2500 |
| DFT(M06-2X) | 6-31G(d,p) | 0.2666 | -0.0167 | 0.2667 | -0.0167 | 0.2667 | -0.0167 | 0.2667 | -0.0167 | 0.2500 | 0.2500 | 0.2500 | 0.2500 |
| DFT(M06-2X) | cc-pVDZ | 0.2642 | -0.0142 | 0.2642 | -0.0142 | 0.2642 | -0.0142 | 0.2642 | -0.0142 | 0.2500 | 0.2500 | 0.2500 | 0.2500 |
| DFT(M06-2X) | cc-pVTZ | 0.2791 | -0.0291 | 0.2791 | -0.0291 | 0.2791 | -0.0291 | 0.2791 | -0.0291 | 0.2500 | 0.2500 | 0.2500 | 0.2500 |
| DFT(M06-2X) | cc-pVQZ | 0.2315 | 0.0185 | 0.2315 | 0.0185 | 0.2315 | 0.0185 | 0.2315 | 0.0185 | 0.2500 | 0.2500 | 0.2500 | 0.2500 |
| DFT(CAM-B3LYP) | 6-31G(d,p) | 0.2652 | -0.0152 | 0.2652 | -0.0152 | 0.2652 | -0.0152 | 0.2652 | -0.0152 | 0.2500 | 0.2500 | 0.2500 | 0.2500 |
| DFT(CAM-B3LYP) | cc-pVDZ | 0.2643 | -0.0143 | 0.2643 | -0.0143 | 0.2643 | -0.0143 | 0.2643 | -0.0143 | 0.2500 | 0.2500 | 0.2500 | 0.2500 |
| DFT(CAM-B3LYP) | cc-pVTZ | 0.2622 | -0.0122 | 0.2622 | -0.0122 | 0.2622 | -0.0122 | 0.2621 | -0.0122 | 0.2500 | 0.2500 | 0.2500 | 0.2500 |
| DFT(CAM-B3LYP) | cc-pVQZ | 0.2791 | -0.0291 | 0.2791 | -0.0291 | 0.2791 | -0.0291 | 0.2791 | -0.0291 | 0.2500 | 0.2500 | 0.2500 | 0.2500 |

**Table S28.** Mulliken atomic spin densities obtained with chosen computational approaches for cationic form of cyclobutadiene. In the right part of the table atomic spin densities of hydrogens are summed into heavy atoms they are connected to. Please note, that data for CASPT2 method are computed for equilibrium geometries obtained at this level of theory, however spin densities are computed basing on CASSCF wavefunction.

# CBDE R+

| Method | Basis set | C1 | H2 | C3 | H4 | C5 | H6 | C7 | H8 | C1+H2 | C3+H4 | C5+H6 | C7+H8 |
|---|---|---|---|---|---|---|---|---|---|---|---|---|---|
| CASPT2, 0-IPEA | 6-31G(d,p) | 0.2489 | 0.0007 | 0.2497 | 0.0007 | 0.2489 | 0.0007 | 0.2497 | 0.0007 | 0.2496 | 0.2504 | 0.2496 | 0.2504 |
| CASPT2, 0-IPEA | cc-pVDZ | 0.2484 | 0.0015 | 0.2485 | 0.0015 | 0.2484 | 0.0015 | 0.2485 | 0.0015 | 0.2499 | 0.2500 | 0.2499 | 0.2500 |
| CASPT2, 0-IPEA | cc-pVTZ | 0.2474 | 0.0028 | 0.2470 | 0.0028 | 0.2474 | 0.0028 | 0.2470 | 0.0028 | 0.2502 | 0.2498 | 0.2502 | 0.2498 |
| CASPT2, S-IPEA | 6-31G(d,p) | 0.2490 | 0.0007 | 0.2496 | 0.0007 | 0.2490 | 0.0007 | 0.2496 | 0.0007 | 0.2497 | 0.2503 | 0.2497 | 0.2503 |
| CASPT2, S-IPEA | cc-pVDZ | 0.2485 | 0.0015 | 0.2485 | 0.0015 | 0.2485 | 0.0015 | 0.2484 | 0.0015 | 0.2500 | 0.2500 | 0.2500 | 0.2499 |
| CASPT2, S-IPEA | cc-pVTZ | 0.2472 | 0.0028 | 0.2472 | 0.0028 | 0.2472 | 0.0028 | 0.2472 | 0.0028 | 0.2500 | 0.2500 | 0.2500 | 0.2500 |
| CASSCF | 6-31G(d,p) | 0.2493 | 0.0008 | 0.2492 | 0.0008 | 0.2493 | 0.0008 | 0.2492 | 0.0008 | 0.2501 | 0.2500 | 0.2501 | 0.2500 |
| CASSCF | cc-pVDZ | 0.2484 | 0.0016 | 0.2484 | 0.0016 | 0.2484 | 0.0016 | 0.2484 | 0.0016 | 0.2500 | 0.2500 | 0.2500 | 0.2500 |
| CASSCF | cc-pVTZ | 0.2472 | 0.0029 | 0.2470 | 0.0029 | 0.2472 | 0.0029 | 0.2470 | 0.0029 | 0.2501 | 0.2499 | 0.2501 | 0.2499 |
| CASSCF | cc-pVQZ | 0.2466 | 0.0034 | 0.2465 | 0.0034 | 0.2466 | 0.0034 | 0.2465 | 0.0034 | 0.2500 | 0.2499 | 0.2500 | 0.2499 |
| DFT(B3LYP) | 6-31G(d,p) | 0.2620 | -0.0120 | 0.2620 | -0.0120 | 0.2620 | -0.0120 | 0.2620 | -0.0120 | 0.2500 | 0.2500 | 0.2500 | 0.2500 |
| DFT(B3LYP) | cc-pVDZ | 0.2622 | -0.0122 | 0.2622 | -0.0122 | 0.2622 | -0.0122 | 0.2622 | -0.0122 | 0.2500 | 0.2500 | 0.2500 | 0.2500 |
| DFT(B3LYP) | cc-pVTZ | 0.2568 | -0.0068 | 0.2568 | -0.0068 | 0.2568 | -0.0068 | 0.2568 | -0.0068 | 0.2500 | 0.2500 | 0.2500 | 0.2500 |
| DFT(B3LYP) | cc-pVQZ | 0.2584 | -0.0084 | 0.2585 | -0.0084 | 0.2584 | -0.0084 | 0.2585 | -0.0084 | 0.2500 | 0.2500 | 0.2500 | 0.2500 |
| DFT(M06-2X) | 6-31G(d,p) | 0.2607 | -0.0107 | 0.2607 | -0.0107 | 0.2607 | -0.0107 | 0.2607 | -0.0107 | 0.2500 | 0.2500 | 0.2500 | 0.2500 |
| DFT(M06-2X) | cc-pVDZ | 0.2606 | -0.0106 | 0.2606 | -0.0106 | 0.2606 | -0.0106 | 0.2606 | -0.0106 | 0.2500 | 0.2500 | 0.2500 | 0.2500 |
| DFT(M06-2X) | cc-pVTZ | 0.2589 | -0.0089 | 0.2589 | -0.0089 | 0.2589 | -0.0089 | 0.2589 | -0.0089 | 0.2500 | 0.2500 | 0.2500 | 0.2500 |
| DFT(M06-2X) | cc-pVQZ | 0.2468 | 0.0032 | 0.2468 | 0.0032 | 0.2468 | 0.0032 | 0.2468 | 0.0032 | 0.2500 | 0.2500 | 0.2500 | 0.2500 |
| DFT(CAM-B3LYP) | 6-31G(d,p) | 0.2618 | -0.0118 | 0.2618 | -0.0118 | 0.2618 | -0.0118 | 0.2618 | -0.0118 | 0.2500 | 0.2500 | 0.2500 | 0.2500 |
| DFT(CAM-B3LYP) | cc-pVDZ | 0.2620 | -0.0120 | 0.2620 | -0.0120 | 0.2619 | -0.0120 | 0.2619 | -0.0120 | 0.2500 | 0.2500 | 0.2500 | 0.2500 |
| DFT(CAM-B3LYP) | cc-pVTZ | 0.2571 | -0.0071 | 0.2571 | -0.0071 | 0.2571 | -0.0071 | 0.2571 | -0.0071 | 0.2500 | 0.2500 | 0.2500 | 0.2500 |
| DFT(CAM-B3LYP) | cc-pVQZ | 0.2595 | -0.0095 | 0.2595 | -0.0095 | 0.2595 | -0.0095 | 0.2595 | -0.0095 | 0.2500 | 0.2500 | 0.2500 | 0.2500 |

Table S29. Mulliken atomic spin densities obtained with chosen computational approaches for the first triplet excited state of cyclobutadiene. In the right part of the table atomic spin densities of hydrogens are summed into heavy atoms they are connected to. Please note, that data for CASPT2 method are computed for equilibrium geometries obtained at this level of theory, however spin densities are computed basing on CASSCF wavefunction.

CBDE R$^0_{T1}$

| Method | Basis set | C1 | H2 | C3 | H4 | C5 | H6 | C7 | H8 | C1+H2 | C3+H4 | C5+H6 | C7+H8 |
|---|---|---|---|---|---|---|---|---|---|---|---|---|---|
| CASPT2, 0-IPEA | 6-31G(d,p) | 0.4981 | 0.0019 | 0.4981 | 0.0019 | 0.4982 | 0.0019 | 0.4981 | 0.0019 | 0.5000 | 0.5000 | 0.5001 | 0.5000 |
| CASPT2, 0-IPEA | cc-pVDZ | 0.4963 | 0.0037 | 0.4962 | 0.0037 | 0.4963 | 0.0037 | 0.4964 | 0.0037 | 0.5000 | 0.4999 | 0.5000 | 0.5001 |
| CASPT2, 0-IPEA | cc-pVTZ | 0.4926 | 0.0074 | 0.4926 | 0.0074 | 0.4926 | 0.0074 | 0.4926 | 0.0074 | 0.5000 | 0.5000 | 0.5000 | 0.5000 |
| CASPT2, S-IPEA | 6-31G(d,p) | 0.4981 | 0.0019 | 0.4981 | 0.0019 | 0.4982 | 0.0019 | 0.4981 | 0.0019 | 0.5000 | 0.5000 | 0.5001 | 0.5000 |
| CASPT2, S-IPEA | cc-pVDZ | 0.4965 | 0.0037 | 0.4960 | 0.0037 | 0.4966 | 0.0037 | 0.4959 | 0.0037 | 0.5002 | 0.4997 | 0.5003 | 0.4996 |
| CASPT2, S-IPEA | cc-pVTZ | 0.4931 | 0.0073 | 0.4922 | 0.0073 | 0.4930 | 0.0074 | 0.4923 | 0.0074 | 0.5004 | 0.4995 | 0.5004 | 0.4997 |
| CASSCF | 6-31G(d,p) | 0.4981 | 0.0019 | 0.4981 | 0.0019 | 0.4981 | 0.0019 | 0.4981 | 0.0019 | 0.5000 | 0.5000 | 0.5000 | 0.5000 |
| CASSCF | cc-pVDZ | 0.4961 | 0.0039 | 0.4961 | 0.0039 | 0.4961 | 0.0039 | 0.4961 | 0.0039 | 0.5000 | 0.5000 | 0.5000 | 0.5000 |
| CASSCF | cc-pVTZ | 0.4925 | 0.0075 | 0.4925 | 0.0075 | 0.4925 | 0.0075 | 0.4925 | 0.0075 | 0.5000 | 0.5000 | 0.5000 | 0.5000 |
| CASSCF | cc-pVQZ | 0.4896 | 0.0104 | 0.4896 | 0.0104 | 0.4896 | 0.0104 | 0.4896 | 0.0104 | 0.5000 | 0.5000 | 0.5000 | 0.5000 |
| DFT(B3LYP) | 6-31G(d,p) | 0.5268 | -0.0268 | 0.5268 | -0.0268 | 0.5268 | -0.0268 | 0.5268 | -0.0268 | 0.5000 | 0.5000 | 0.5000 | 0.5000 |
| DFT(B3LYP) | cc-pVDZ | 0.5260 | -0.0260 | 0.5260 | -0.0260 | 0.5260 | -0.0260 | 0.5260 | -0.0260 | 0.5000 | 0.5000 | 0.5000 | 0.5000 |
| DFT(B3LYP) | cc-pVTZ | 0.5172 | -0.0172 | 0.5172 | -0.0172 | 0.5172 | -0.0172 | 0.5172 | -0.0172 | 0.5000 | 0.5000 | 0.5000 | 0.5000 |
| DFT(B3LYP) | cc-pVQZ | 0.5186 | -0.0186 | 0.5186 | -0.0186 | 0.5186 | -0.0186 | 0.5186 | -0.0186 | 0.5000 | 0.5000 | 0.5000 | 0.5000 |
| DFT(M06-2X) | 6-31G(d,p) | 0.5279 | -0.0279 | 0.5279 | -0.0279 | 0.5279 | -0.0279 | 0.5279 | -0.0279 | 0.5000 | 0.5000 | 0.5000 | 0.5000 |
| DFT(M06-2X) | cc-pVDZ | 0.5246 | -0.0246 | 0.5246 | -0.0246 | 0.5246 | -0.0246 | 0.5246 | -0.0246 | 0.5000 | 0.5000 | 0.5000 | 0.5000 |
| DFT(M06-2X) | cc-pVTZ | 0.5330 | -0.0330 | 0.5330 | -0.0330 | 0.5330 | -0.0330 | 0.5330 | -0.0330 | 0.5000 | 0.5000 | 0.5000 | 0.5000 |
| DFT(M06-2X) | cc-pVQZ | 0.4744 | 0.0256 | 0.4744 | 0.0256 | 0.4744 | 0.0256 | 0.4744 | 0.0256 | 0.5000 | 0.5000 | 0.5000 | 0.5000 |
| DFT(CAM-B3LYP) | 6-31G(d,p) | 0.5265 | -0.0265 | 0.5265 | -0.0265 | 0.5265 | -0.0265 | 0.5265 | -0.0265 | 0.5000 | 0.5000 | 0.5000 | 0.5000 |
| DFT(CAM-B3LYP) | cc-pVDZ | 0.5256 | -0.0256 | 0.5256 | -0.0256 | 0.5256 | -0.0256 | 0.5256 | -0.0256 | 0.5000 | 0.5000 | 0.5000 | 0.5000 |
| DFT(CAM-B3LYP) | cc-pVTZ | 0.5181 | -0.0181 | 0.5181 | -0.0181 | 0.5181 | -0.0181 | 0.5181 | -0.0181 | 0.5000 | 0.5000 | 0.5000 | 0.5000 |
| DFT(CAM-B3LYP) | cc-pVQZ | 0.5218 | -0.0218 | 0.5218 | -0.0218 | 0.5218 | -0.0218 | 0.5218 | -0.0218 | 0.5000 | 0.5000 | 0.5000 | 0.5000 |

Table S30. Mulliken atomic spin densities for cyclobutadiene in the ground electronic state predicted as combination ($R^- + R^+ - R^0_{T1}$) of atomic spin densities computed for other electronic states structures. In the right part of the table there are data for summed spin densities of heavy atoms and hydrogens connected to them. Please note, that data for CASPT2 method are computed for equilibrium geometries obtained at this level of theory, however spin densities are computed basing on CASSCF wavefunction.

CBDE $R^- + R^+ - R^0_{T1}$

| Method | Basis set | C1 | H2 | C3 | H4 | C5 | H6 | C7 | H8 | C1+H2 | C3+H4 | C5+H6 | C7+H8 |
|---|---|---|---|---|---|---|---|---|---|---|---|---|---|
| CASPT2, 0-IPEA | 6-31G(d,p) | 0.0019 | -0.0001 | -0.0016 | -0.0001 | 0.0018 | -0.0001 | -0.0016 | -0.0001 | 0.0018 | -0.0017 | 0.0017 | -0.0017 |
| CASPT2, 0-IPEA | cc-pVDZ | 0.0014 | 0.0000 | -0.0012 | -0.0001 | 0.0013 | 0.0000 | -0.0014 | -0.0001 | 0.0014 | -0.0013 | 0.0013 | -0.0015 |
| CASPT2, 0-IPEA | cc-pVTZ | -0.0012 | 0.0003 | 0.0006 | 0.0003 | -0.0012 | 0.0003 | 0.0006 | 0.0003 | -0.0009 | 0.0009 | -0.0009 | 0.0009 |
| CASPT2, S-IPEA | 6-31G(d,p) | -0.0001 | -0.0001 | 0.0003 | -0.0001 | -0.0002 | -0.0001 | 0.0003 | -0.0001 | -0.0002 | 0.0002 | -0.0003 | 0.0002 |
| CASPT2, S-IPEA | cc-pVDZ | 0.0004 | 0.0000 | -0.0002 | 0.0000 | 0.0003 | 0.0000 | -0.0002 | 0.0000 | 0.0004 | -0.0002 | 0.0003 | -0.0002 |
| CASPT2, S-IPEA | cc-pVTZ | -0.0004 | 0.0004 | -0.0003 | 0.0004 | -0.0005 | 0.0003 | -0.0003 | 0.0003 | 0.0000 | 0.0001 | -0.0002 | 0.0000 |
| CASSCF | 6-31G(d,p) | 0.0001 | 0.0000 | 0.0000 | 0.0000 | 0.0001 | 0.0000 | 0.0000 | 0.0000 | 0.0001 | 0.0000 | 0.0001 | 0.0000 |
| CASSCF | cc-pVDZ | 0.0001 | -0.0001 | 0.0001 | -0.0001 | 0.0001 | -0.0001 | 0.0001 | -0.0001 | 0.0000 | 0.0000 | 0.0000 | 0.0000 |
| CASSCF | cc-pVTZ | -0.0005 | 0.0004 | -0.0003 | 0.0004 | -0.0005 | 0.0004 | -0.0003 | 0.0004 | -0.0001 | 0.0001 | -0.0001 | 0.0001 |
| CASSCF | cc-pVQZ | -0.0028 | 0.0027 | -0.0026 | 0.0027 | -0.0028 | 0.0027 | -0.0026 | 0.0027 | -0.0001 | 0.0001 | -0.0001 | 0.0001 |
| DFT(B3LYP) | 6-31G(d,p) | 0.0004 | -0.0004 | 0.0004 | -0.0004 | 0.0004 | -0.0004 | 0.0004 | -0.0004 | 0.0000 | 0.0000 | 0.0000 | 0.0000 |
| DFT(B3LYP) | cc-pVDZ | 0.0006 | -0.0006 | 0.0006 | -0.0006 | 0.0006 | -0.0006 | 0.0006 | -0.0006 | 0.0000 | 0.0000 | 0.0000 | 0.0000 |
| DFT(B3LYP) | cc-pVTZ | 0.0009 | -0.0009 | 0.0009 | -0.0009 | 0.0009 | -0.0009 | 0.0009 | -0.0009 | 0.0000 | 0.0000 | 0.0000 | 0.0000 |
| DFT(B3LYP) | cc-pVQZ | 0.0003 | -0.0003 | 0.0004 | -0.0003 | 0.0003 | -0.0003 | 0.0003 | -0.0003 | 0.0000 | 0.0000 | 0.0000 | 0.0000 |
| DFT(M06-2X) | 6-31G(d,p) | -0.0005 | 0.0005 | -0.0005 | 0.0005 | -0.0006 | 0.0005 | -0.0005 | 0.0005 | 0.0000 | 0.0000 | 0.0000 | 0.0000 |
| DFT(M06-2X) | cc-pVDZ | 0.0002 | -0.0002 | 0.0002 | -0.0002 | 0.0001 | -0.0002 | 0.0002 | -0.0002 | 0.0000 | 0.0000 | 0.0000 | 0.0000 |
| DFT(M06-2X) | cc-pVTZ | 0.0050 | -0.0050 | 0.0050 | -0.0050 | 0.0050 | -0.0050 | 0.0050 | -0.0050 | 0.0000 | 0.0000 | 0.0000 | 0.0000 |
| DFT(M06-2X) | cc-pVQZ | 0.0039 | -0.0039 | 0.0040 | -0.0039 | 0.0039 | -0.0039 | 0.0039 | -0.0039 | 0.0000 | 0.0000 | 0.0000 | 0.0000 |
| DFT(CAM-B3LYP) | 6-31G(d,p) | 0.0005 | -0.0005 | 0.0004 | -0.0005 | 0.0005 | -0.0005 | 0.0005 | -0.0005 | 0.0000 | 0.0000 | 0.0000 | 0.0000 |
| DFT(CAM-B3LYP) | cc-pVDZ | 0.0007 | -0.0006 | 0.0007 | -0.0006 | 0.0006 | -0.0006 | 0.0006 | -0.0006 | 0.0000 | 0.0000 | 0.0000 | 0.0000 |
| DFT(CAM-B3LYP) | cc-pVTZ | 0.0012 | -0.0011 | 0.0011 | -0.0011 | 0.0011 | -0.0011 | 0.0011 | -0.0011 | 0.0001 | 0.0000 | 0.0000 | -0.0001 |
| DFT(CAM-B3LYP) | cc-pVQZ | 0.0169 | -0.0168 | 0.0168 | -0.0168 | 0.0168 | -0.0168 | 0.0168 | -0.0168 | 0.0000 | 0.0000 | 0.0000 | 0.0000 |

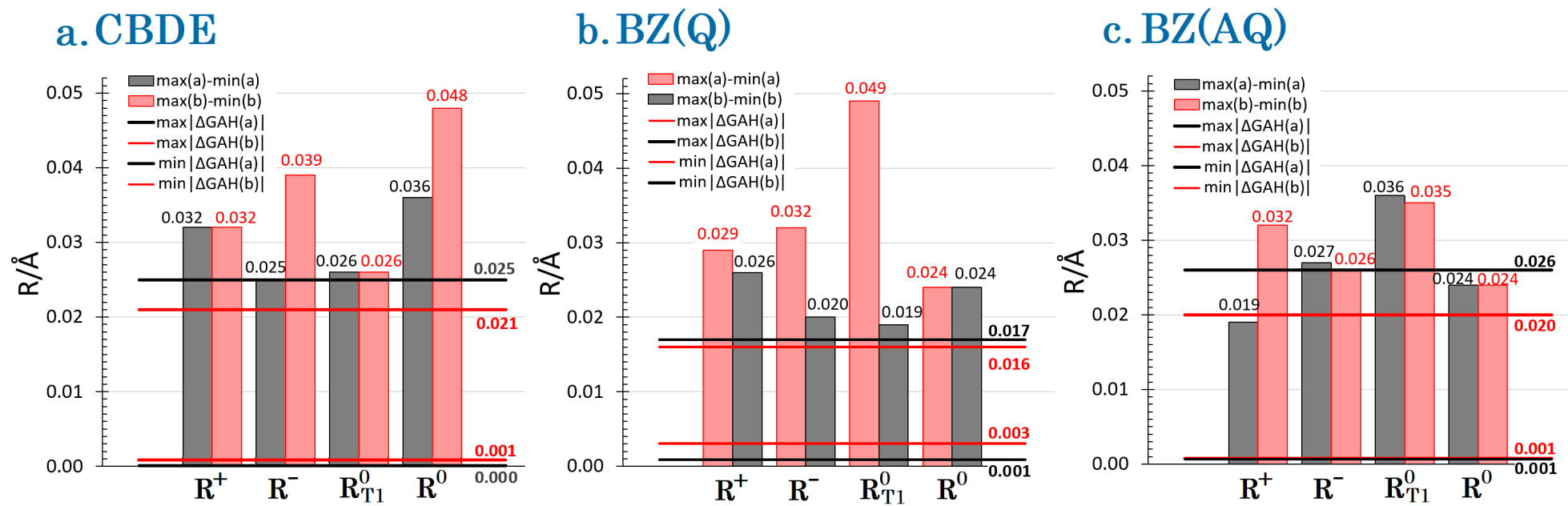

**Figure S1.** Minimum (**min|ΔGAH(R)|**) and maximum (**max|ΔGAH(R)|**) of unsigned values of **Eq. 1** expression on the background of statistic ranges **max(R)-min(R)** of **a** and **b** bonds lengths of quinoid (**Q**) and anti-quinoid (**AQ**) benzene (**BZ**) conformers and for cyclobutadiene (**CBDE**) for all investigated computational approaches, and all basis sets applied. Data for shorter bond are marked as red of pink, whereas data for longer ones are black or grey (see **Fig. 1**). For more detailed data see **Tables S1-S14**.